\documentclass[11pt, A4]{article}

\usepackage{graphicx}
\usepackage{epstopdf}
\usepackage{amsmath}
\usepackage{amssymb}
\usepackage{amsfonts}
\usepackage{amsthm}
\usepackage[usenames]{color}
\usepackage{array}
\usepackage{tabularx}
\usepackage{booktabs}           
\usepackage{caption}
\usepackage{multirow}

\usepackage[
      colorlinks=true,
      linkcolor=blue,
      urlcolor=blue,
      filecolor=blue,
      citecolor=red,
      pdfstartview=FitV,
      pdftitle={},
      pdfauthor={},
      pdfsubject={},
      pdfkeywords={},
      pdfpagemode=None,
      bookmarksopen=true
]{hyperref}

\usepackage{empheq}
\usepackage{mathtools}
\usepackage{epsfig}

\usepackage{hyperref}

\def\beq{\begin{eqnarray}}
\def\eeq{\end{eqnarray}}

\def\mf{\mathfrak}

\def\a{\alpha}
\def\b{\beta}

\def\e{\epsilon}

\def\S{\Sigma}

\def\x{\xi}

\def\be{\begin{equation}}
\def\ee{\end{equation}}
\def\bea{\begin{eqnarray}}
\def\eea{\end{eqnarray}}

\newcommand{\rom}[1]{\mathrm{#1}}

\def\so{\mathfrak{so}}

\def\cL{\mathcal{L}}
\def\cM{\mathcal{M}}

\def\cQ{\mathcal{Q}}

\def\cV{\mathcal{V}}

\def\mf{\mathfrak}
\def\nn{\nonumber}

\usepackage{dynkin-diagrams}

\numberwithin{equation}{section}

\renewcommand{\thefootnote}{\fnsymbol{footnote}}

\begin{document}

\begin{centering}

\thispagestyle{empty}

{\flushright {Preprint CMI 2018}\\[15mm]}

{\LARGE \textsc{Geroch Group Description \\  \vskip 0.2cm of Bubbling Geometries}} \\

 \vspace{0.8cm}

{\large 
Pratik Roy$^{1}$ and  Amitabh Virmani$^{1,2,3, }$}\footnote[1]{Currently on lien from Institute of Physics, Sachivalaya Marg, Bhubaneswar, Odisha, India 751005. }
\vspace{0.5cm}

\begin{minipage}{.9\textwidth}\small  \begin{center}
$^1$Chennai Mathematical Institute, H1 SIPCOT IT Park, \\ Kelambakkam, Tamil Nadu, India 603103\\
  \vspace{0.5cm}
$^2$Institute of Physics, Sachivalaya Marg, \\ Bhubaneswar, Odisha, India 751005 \\
  \vspace{0.5cm}
$^3$Homi Bhabha National Institute, Training School Complex, \\ Anushakti Nagar, Mumbai 400085, India \\
  \vspace{0.5cm}
{\tt proy, avirmani@cmi.ac.in}
\\ $ \, $ \\

\end{center}
\end{minipage}

\end{centering}

\renewcommand{\thefootnote}{\arabic{footnote}}

\begin{abstract}
The Riemann-Hilbert approach to studying solutions of supergravity theories allows us to associate spacetime independent monodromy matrices (matrices in the Geroch group) with  solutions that effectively only depend on two spacetime coordinates. This offers insights into symmetries of supergravity theories, and in the classification of their solutions.  In this paper, we initiate a systematic study of monodromy matrices for multi-center solutions of five-dimensional U(1)$^3$ supergravity.  We obtain monodromy matrices for a class of collinear Bena-Warner bubbling geometries. We show that for this class of solutions, monodromy matrices in the vector representation of SO(4,4) have only simple poles with residues of rank two and nilpotency degree two.  These properties strongly suggest that an inverse scattering construction along the lines of [arXiv:1311.7018 [hep-th]] can be given for this class of solutions, though it is not attempted in this work. Along the way, we clarify a technical point in the existing literature:  we show that the so-called ``spectral flow transformations'' of Bena, Bobev, and Warner are precisely a class of Harrison transformations when restricted to the situation of two commuting Killing symmetries in five-dimensions. 
\end{abstract}

\newpage
\tableofcontents

\setcounter{equation}{0}

\section{Introduction}

The Riemann-Hilbert approach \cite{BM, BMnotes} to studying solutions to supergravity theories is remarkable in that it allows to study solutions that effectively only depend on two spacetime coordinates in terms of spacetime independent monodromy matrices. More precisely, on the one hand, to a  given solution one can associate a monodromy matrix, on the other hand, given a candidate monodromy matrix one can perform its canonical factorisation with prescribed analyticity properties  to  obtain explicit solutions of  supergravity theories.

Thus, in this approach, the problem of solving non-linear partial differential equations to obtain solutions of supergravity, is mapped into a matrix valued factorisation problem in one complex variable.  This approach offers significant insight into duality symmetries of  supergravity theories, and also into the organisation and classification of its solutions.  It is closely related to the so-called inverse scattering approach \cite{BZ1,BZ2,BV} that has been astonishingly successful for understanding solutions of five-dimensional vacuum gravity \cite{Emparan,Iguchi, Rocha:2013qya}.

In order to obtain explicit solutions in the Riemann-Hilbert approach a canonical factorisation must be performed. Over the last few years at least two different approaches have been developed. In the first approach  \cite{KKV1, KKV2, KKV3,Chakrabarty:2014ora}, the authors have focused on monodromy matrices with simple poles with suitable rank residues. In the the second approach \cite{Camara:2017hez, Cardoso:2017cgi}, the authors have converted the matrix valued factorisation problem into a vectorial Riemann-Hilbert problem and solved it using complex analysis. Several examples have been worked out in both these approaches. A construction of the JMaRT \cite{Jejjala:2005yu} solution was worked out in \cite{KKV3}, and its relation to the Belinsky-Zakharov inverse scattering construction was explored in  \cite{Chakrabarty:2016nbu}.

 In this paper we initiate a systematic study of monodromy matrices for multi-center solutions.  The main motivation for this study is as follows. A generalisation of various known multi-center solutions to non-supersymmetric setting is a problem that has received much attention in recent years 
 \cite{Bossard:2014yta, Bossard:2014ola, Bena:2015drs, Bena:2016dbw, Bossard:2017vii}, and a variety of solutions have been obtained.  Alternative approaches to such solutions, together with developments of different techniques, are much desirable to understand better the spectrum and dynamics of such solutions. In this paper we take first steps in this direction.
  
We obtain monodromy matrices for a class of collinear Bena-Warner bubbling geometries \cite{Bena:2005va, Berglund:2005vb}. We consider these solutions as embedded in five-dimensional U(1)$^3$ supergravity. The five-dimensional  U(1)$^3$ supergravity has SO(4,4) hidden symmetry when reduced to three-dimensions, and it has affine SO(4,4) (an infinite dimensional symmetry) as its two-dimensional duality symmetry group. This affine symmetry is called the Geroch group. The monodromy matrices are matices in this group: $8 \times 8$ matrices of one complex variable in the defining representation of SO(4,4). 

We show that for collinear Bena-Warner bubbling  solutions, monodromy matrices have only simple poles with residues of rank two and nilpotency degree two.  These are precisely some of the conditions required for the Riemann-Hilbert factorisation developed in \cite{KKV2}. 
We have not explored the explicit factorisation of monodromy matrices in this work; we leave this investigation for future studies.

The rest of the paper is organised as follows. In section \ref{sec:dimred} we present 
an appropriate dimensional reduction of five-dimensional U(1)$^3$ supergravity to three dimensions. This dimensional reduction is well adapted to obtain  a coset description of the Bena-Warner solutions, which is worked out in detail in section
\ref{sec:BW_coset}.   The Bena-Warner class of solutions have two well studied symmetries called ``gauge transformations'' and ``spectral flow transformations'' \cite{Bena:2008wt}. Under these  transformations solutions are mapped to solutions within the Bena-Warner class. We show that from the three-dimensional duality symmetry point of view, the so-called ``gauge transformations'' are simple shifts of certain three-dimensional scalars, and the so-called ``spectral flow transformations'' are a class of Harrison transformations. 

In section \ref{sec:Geroch} we obtain the Geroch group (monodromy) matrices for collinear multi-center bubbling solutions. In section \ref{sec:examples} we workout a few illustrative but non-trivial examples. We end with a brief summary and possible future directions in section \ref{sec:disc}. 

In  appendix
\ref{app:nilpotent_orbits}
 we work out explicit representatives for smaller nilpotent orbits of the $\mathfrak{so}(4,4)$ Lie algebra. We  list matrix rank and nilpotency degree of these representatives in the defining representation of $\mathfrak{so}(4,4)$.

\section{Dimensional reduction}
\label{sec:dimred}
We start by presenting an appropriate dimensional reduction of five-dimensional U(1)$^3$ supergravity to three dimensions. 
It is well known that the U(1)$^3$ supergravity when dimensionally reduced to three-dimensions has an SO(4,4)  symmetry. In this section we make this manifest, and the notation introduced in the discussion will be used through the rest of the paper. 
The sign conventions below are same as \cite{Sahay:2013xda} except for the over-all sign of the Chern-Simons term in the 11d action; we refer the reader to that reference for further details and references. This sign difference results in different signs compared to that reference in some equations.

Our starting point is the Lagrangian of eleven-dimensional supergravity,
\be
\mathcal{L}_{11} = R_{11} \star_{11} \mathbf{1}  - \frac{1}{2} F_{[4]} \wedge \star_{11} F_{[4]}
- \frac{1}{6} F_{[4]} \wedge F_{[4]} \wedge A_{[3]}.
\ee
We are interested in the five-dimensional theory obtained by dimensional reduction on T$^6$ with the following metric ansatz,
\be
ds^2_{11} = ds^2_{5} + h^1 (d\tilde x_1^2 + d\tilde x_2^2) + h^2 (d\tilde x_3^2 + d\tilde x_4^2)+h^3 (d\tilde x_5^2 + d\tilde x_6^2),
\ee 
together with the form-field ansatz,
\be
A_{[3]} = A^1_{[1]} \wedge d\tilde x_1  \wedge d\tilde x_2  +A^2_{[1]} \wedge d\tilde x_3  \wedge d\tilde x_4 +A^3_{[1]} \wedge d\tilde x_5  \wedge d\tilde x_6.
\ee
We work with the assumption that nothing depends on the six torus coordinates $\tilde x_i$. We also assume that the scalars $h^I$, with $I = 1,2,3,$ obey the constraint\footnote{For a discussion on the constraint \eqref{constraint} see e.g.~\cite{Colgain:2014pha}. We thank E.~Colg\'ain for bringing this reference to our attention. }
\be
h^1 h^2 h^3 = 1, \label{constraint}
\ee
i.e., we assume that all complex structure moduli of the T$^6$ as well as the volume modulus are frozen. The constraint 
\eqref{constraint} ensures that the resulting five-dimensional metric is in the Einstein frame.
The eleven dimensional Lagrangian
reduces to the five-dimensional U(1)$^3$ supergravity,
\begin{equation}
\label{eqn:Lagrangian5d}
\mathcal{L}_5 = R_5 \star_5 \mathbf{1}  - \frac{1}{2} G_{IJ} \star_5 dh^I \wedge  dh^J
- \frac{1}{2}G_{IJ}\star_5 F^I_{[2]} \wedge F^J_{[2]}
- \frac{1}{6} C_{IJK} F^I_{[2]} \wedge F^J_{[2]} \wedge A^K_{[1]},
\end{equation}
where $C_{IJK} = |\epsilon_{IJK}|$, and $G_{IJ}$ is diagonal with entries $G_{II}
= (h^I)^{-2}$. Note that the constraint \eqref{constraint} must be solved before computing variations of the action in order
to obtain equations of motion for various fields. It can be solved, say, with the following choice,
\be
h^1 = e^{\sqrt{\frac{2}{3}}\Psi}, \quad h^2 = e^{- \sqrt{\frac{1}{6}}\Psi - \sqrt{\frac{1}{2}}\Phi}, \quad h^3 = e^{- \sqrt{\frac{1}{6}}\Psi + \sqrt{\frac{1}{2}}\Phi}.
\ee

\subsection{Timelike reduction: 5d to 4d}
To perform Kaluza-Klein reduction from five-dimensions to four-dimensions we parameterize our 5d spacetime as
\be
\label{an1}
ds^2 = \epsilon_1 f^2  (dt + \check A^0_{[1]})^2 + f^{-1}ds^2_4,
\ee
and 5d vectors as
\be
\label{an2}
A^I_{[1]} = \chi^{I}(dt + \check A^0_{[1]}) + \check A^I_{[1]},
\ee
where we use $\epsilon_1$ to keep track of minus signs.  The case of interest for the present discussion is $\epsilon_1 = -1$. The timelike reduction is thought of as an effective simplification of the five-dimensional dynamics in the presence of $\partial_t$ Killing vector.

The 4d graviphoton $\check A^0_{[1]}$ and the 4d vectors $\check A^I_{[1]}$ form a symplectic vector $\check A^\Lambda_{[1]}$ with $\Lambda = 0,1,2,3$. We define the field strength for these vectors as simply $F_{[2]}^\Lambda = d \check A_{[1]}^\Lambda$.
Inserting ansatzes \eqref{an1} and \eqref{an2} in Lagrangian \eqref{eqn:Lagrangian5d}  we obtain,
\bea
\mathcal{L}_4 &=&  R \star_4 \mathbf{1} - \frac{1}{2}G_{IJ} \star_4 d h^I \wedge
d h^J  - \frac{3}{2f^2}\star_4 df \wedge df   - \epsilon_1 \frac{f^3}{2}\star_4
\check F_{[2]}^0 \wedge \check F_{[2]}^0 \nn  \\
&& - \epsilon_1 \frac{1}{2 f^2}G_{IJ} \star_4 d \chi^I \wedge d \chi^J  - \frac{f}{2}
G_{IJ} \star_4 (\check F_{[2]}^I +\chi^I \check F_{[2]}^0) \wedge (\check
F_{[2]}^J +\chi^J \check F_{[2]}^0) \nn
\\
&&-\frac{1}{2}C_{IJK}\chi^I \check F_{[2]}^J  \wedge \check F_{[2]}^K  -
\frac{1}{2} C_{IJK}\chi^I \chi^J \check F_{[2]}^0  \wedge \check F_{[2]}^K - \frac{1}{6}
C_{IJK} \chi^I \chi^J \chi^K \check F_{[2]}^0 \wedge \check F_{[2]}^0\, .
 \label{STUaction}
\eea
Note that the sign of the kinetic term for the graviphoton $\check F_{[2]}^0$ and that for the scalars $\chi^I$ depend on the sign $\epsilon_1$.

The reduced Lagrangian \eqref{STUaction} can be obtained from a cubic prepotential using split complex numbers, where the  imaginary unit $e$ squares to $+1$ instead of $-1$,
$
 \bar{e} = - e, \: e^2 = +1.
$
Specifically, taking 
\be
F(X) = - \frac{X^1 X^2 X^3}{X^0},
\ee
and using the gauge  $X^0 = 1$ and $X^I = - \chi^I + e f h^I$, Lagrangian \eqref{STUaction} with $\epsilon_1 = -1$ is seen to be identical to
\be
{\cal L}_4 = R \star_4 \mathbf{1} - 2 g_{I \bar{J}} \star_4 d X^I \wedge d \bar{X}^{\bar{J}} + \frac{1}{2} \check F^\Lambda_{[2]} \wedge \check G_{\Lambda [2]},
\ee
where $\check F^\Lambda_{[2]} = d \check A^\Lambda_{[1]}$. The indices $I, J$ run from 1 to $3$, and $g_{I \bar{J}} = \partial_I \partial_{\bar{J}} K$ with the potential
\be
K = - \log \left[ - e (\bar{X}^\Lambda F_\Lambda - \bar{F}_\Lambda X^\Lambda)\right],
\ee
and where $F_\Lambda = \partial_{\Lambda} F$. The two form  $\check G_{\Lambda [2]}$ is defined
as
\be
\check G_{\Lambda [2]} = - (\mathrm{Re} N)_{\Lambda \Sigma} \check F^{\Sigma}_{[2]} + (\mathrm{Im} N)_{\Lambda \Sigma} \star_4 \check F^{\Sigma}_{[2]},
\ee
where the split complex symmetric matrix $N_{\Lambda \Sigma}$ is constructed from the prepotential as
\be
N_{\Lambda \Sigma} = \bar{F}_{\Lambda \Sigma} + 2 e \frac{(\mathrm{Im} F \cdot X)_\Lambda (\mathrm{Im} F \cdot X)_\Sigma}{X \cdot\mathrm{Im} F \cdot X },
\ee
and where $F_{\Lambda \Sigma} = \partial_\Lambda \partial_\Sigma F$. 
Explicitly, the $\mbox{Re} N$ and $\mbox{Im} N$ matrices are as follows
\be
\mbox{Re} N = \left(
\begin{array}{cccc}
 2 \chi_1 \chi_2 \chi_3 & \chi_2 \chi_3 & \chi_1 \chi_3 & \chi_1
   \chi_2 \\
 \chi_2 \chi_3 & 0 & \chi_3 & \chi_2 \\
 \chi_1 \chi_3 & \chi_3 & 0 & \chi_1 \\
 \chi_1 \chi_2 & \chi_2 & \chi_1 & 0 \\
\end{array}
\right),
\ee 
\be
\mbox{Im} N= f \left(
\begin{array}{cccc}
 f^2+\sum_{i=1}^{3}\frac{\chi_i^2}{(h^i)^2} & -\frac{\chi_1}{(h^1)^2} &
   -\frac{\chi_2}{(h^2)^2} & -\frac{\chi_3}{(h^3)^2} \\
 -\frac{\chi_1}{(h^1)^2} & -\frac{1}{(h^1)^2} & 0 & 0 \\
 -\frac{\chi_2}{(h^2)^2} & 0 & -\frac{1}{(h^2)^2} & 0 \\
 -\frac{\chi_3}{(h^3)^2} & 0 & 0 & -\frac{1}{(h^3)^2} \\
\end{array}
\right).
\ee
\subsection{Spacelike reduction: 4d to 3d}
Now we perform a spacelike reduction from four to three dimensions. For this  we parameterize our four-dimensional space as
\be
ds^2_4 = e^{2U} (dz + \omega_3)^2 + e^{-2 U}ds^2_3, \label{an3}
\ee
and the 4d one-forms as
\be
\check A^\Lambda_{[1]} = \zeta^\Lambda (dz + \omega_3) + A_3^\Lambda, \label{an4}
\ee
where $A_3^\Lambda$ and $\omega_3$ are three-dimensional one-forms. 
We define the field strengths simply as $F_3^\Lambda := d A_3^\Lambda$ and $F_3 := d \omega_3$. 
Since in three dimensions, vector fields are dual to scalars, we now dualise  vectors $A_3^\Lambda$ and $\omega_3$ into scalars $\tilde \zeta_\Lambda$ and $\sigma$.
The procedure of dualisation interchanges the role of Bianchi identities and field equations. The easiest way to achieve dualisation is to treat $F_3^\Lambda$ and $F_3$ as fundamental fields in their own right and impose Bianchi identities through Lagrange multipliers. To this end we add the following Lagrange multiplier terms to the 3d Lagrangian
\be
 +\tilde \zeta_\Lambda F_3^\Lambda + \frac{1}{2}(\sigma + \zeta^\Lambda \tilde \zeta_\Lambda) F_3.
\ee
Thus the total three dimensional Lagrangian we consider is
\bea
{\cal L}_3 &=& R \star_3 \mathbf{1} - 2 \star_3 d U \wedge dU - \frac{1}{2} e^{4U} \star_3 F_3 \wedge F_3  - 2 g_{I\bar{J}} \star_3 d X^{I} \wedge d \bar{X}^{\bar{J}} \nn \\
 & & + \frac{1}{2} e^{2U}(\mathrm{Im}N)_{\Lambda \Sigma} \star_3 (F_3^\Lambda + \zeta^\Lambda F_3)\wedge (F_3^\Sigma + \zeta^\Sigma F_3) + \frac{1}{2} e^{-2U} (\mathrm{Im}N)_{\Lambda \Sigma}\star_3 d \zeta^\Lambda \wedge d \zeta^\Sigma \nn \\
 & & - (\mathrm{Re}N)_{\Lambda \Sigma} \: d \zeta^\Lambda \wedge (F_3^\Sigma + \zeta^\Sigma F_3)
 + \tilde \zeta_\Lambda d F_3^\Lambda + \frac{1}{2}(\sigma + \zeta^\Lambda \tilde \zeta_\Lambda) d F_3.
 \label{Lag3d1}
\eea
Clearly, variations of this Lagrangian with respect to $\sigma$ and $\tilde \zeta_\Lambda$ give the required Bianchi identities.  Upon integration by parts on the Lagrange multiplier terms, equations for $F_3^\Sigma$ and $F_3$ are purely algebraic. These equations allow us to do the dualizations of the one-forms. We find
\be
\label{dual1}
 d \tilde \zeta_\Lambda = e^{2U}(\mathrm{Im}N)_{\Lambda \Sigma} \star_3 (F_3^\Sigma + \zeta^\Sigma F_3) - (\mathrm{Re}N)_{\Lambda \Sigma} d\zeta^\Sigma,
\ee
and
\be
\label{dual2}
 d \sigma  + \tilde \zeta_\Lambda d \zeta^\Lambda   - \zeta^\Lambda d \tilde\zeta_\Lambda   = - 2 e^{4U} \star_3 F_3.
\ee
Substituting these back into Lagrangian \eqref{Lag3d1} we find that it takes the form
\bea
\cL_3 = R \star_3 \mathbf{1} -\frac{1}{2} G_{ab} d \varphi^a \wedge \star d
\varphi^b.
\eea
Metric $G_{ab}$ in our conventions
 is
\bea
 G_{ab}d\varphi^a d\varphi^b &=& 4 dU^2 + 4 g_{I \bar{J}}dX^I d\bar{X}^{\bar J} -
 \frac{1}{4}e^{-4U} \left( d\sigma +  \tilde \zeta_\Lambda d \zeta^\Lambda -
 \zeta^\Lambda d \tilde \zeta_\Lambda \right)^2 \label{cmapst} \\
 && \hspace{-2cm} + e^{-2U}\left[ -(\mbox{Im} N)_{\Lambda \Sigma}d\zeta^\Lambda
 d\zeta^\Sigma + ((\mbox{Im} N)^{-1})^{\Lambda \Sigma} \left( d\tilde
 \zeta_\Lambda +(\mbox{Re} N)_{\Lambda \Xi} d\zeta^\Xi \right)  \left( d\tilde
 \zeta_\Sigma +(\mbox{Re} N)_{\Sigma \Gamma} d\zeta^\Gamma \right) \nonumber \right].
\eea
This metric is identical to the one obtained in \cite{Sahay:2013xda}.

The symmetric space \eqref{cmapst} can  be parameterized in the Iwasawa gauge by the coset
element \cite{Bossard:2009we}
\be
\label{iwasawa}
\cV = e^{- U \, \mathbb{H}_0} \cdot \prod_{I=1,2,3}\left(e^{-\frac{1}{2} \left[\log (f h^I)\right] \mathbb{H}_I} \cdot e^{ \chi^I \mathbb{E}_I} \right) \cdot
e^{-\zeta^\Lambda \mathbb{E}_{q_\Lambda}-  \tilde \zeta_\Lambda \mathbb{E}_{p^\Lambda}}\cdot
e^{-\frac{1}{2}\sigma \mathbb{E}_0},
 \ee
where for an  explicit parametrisation of the Lie algebra 
$\mathfrak{so}(4,4)\ni \mathbb{X}=\sum {\underline E}_a {\mathbb{E}_a}$ we use
\be
\label{Xso44}
\mathbb{X}=
\left(
\begin{array}{cccccccc}
 {\underline H}_{2}+{\underline H}_{3} & -{\underline E}_{3} & -{{\underline F}_{q_1}}& {{\underline F}_{q_0}}& 0 & -{\underline E}_{2} &
   {{\underline E}_{p^0}}& {{\underline E}_{p^1}}\\
 -{\underline F}_{3} & {\underline H}_{2}-{\underline H}_{3} & -{{\underline F}_{p^2}}& {{\underline F}_{q_3}}& {\underline E}_{2} & 0 &
   {{\underline E}_{p^3}}& -{{\underline E}_{q_2}}\\
 -{{\underline E}_{q_1}}& -{{\underline E}_{p^2}}& {\underline H}+{\underline H}_{1} & -{\underline E}_{1} & -{{\underline E}_{p^0}}&
   -{{\underline E}_{p^3}}& 0 & -{\underline E}_0 \\
 {{\underline E}_{q_0}}& {{\underline E}_{q_3}}& -{\underline F}_{1} & {\underline H}-{\underline H}_{1} & -{{\underline E}_{p^1}}&
   {{\underline E}_{q_2}}& {\underline E}_0 & 0 \\
 0 & {\underline F}_{2} & -{{\underline F}_{p^0}}& -{\underline F}_{p^1}
 & -{\underline H}_{2}-{\underline H}_{3} & {\underline F}_{3} &
   {{\underline E}_{q_1}}& -{{\underline E}_{q_0}}\\
 -{\underline F}_{2} & 0 & -{{\underline F}_{p^3}}& {{\underline F}_{q_2}}& {\underline E}_{3} & {\underline H}_{3}-{\underline H}_{2} &
   {{\underline E}_{p^2}}& -{{\underline E}_{q_3}}\\
 {{\underline F}_{p^0}}& {{\underline F}_{p^3}}& 0 & {\underline F}_{0} & {{\underline F}_{q_1}}&
   {{\underline F}_{p^2}}& -{\underline H}-{\underline H}_{1} & {\underline F}_{1} \\
 {\underline F}_{p^1}& -{{\underline F}_{q_2}}& -{\underline F}_{0} & 0 & -{{\underline F}_{q_0}}&
   -{{\underline F}_{q_3}}& {\underline E}_{1} & {\underline H}_{1}-{\underline H}
\end{array}
\right). \nn
\ee
In the above equation ${\mathbb{E}_a}$ are the 28 generators 
and ${\underline E}_a$ are the 28 dual coordinates. The generators ${\mathbb{E}_a}$ can be readily written in the basis \eqref{basis1}--\eqref{basis3}\footnote{Since we are doing dimensional reduction first over time  and then over space, the names of the generators $\mathbb{E}_{q_\Lambda}$ and  $\mathbb{E}_{p^\Lambda}$ do not have the same interpretation as they have in reference \cite{Bossard:2009we}. However, since in many papers this notation is used, we continue to use it for writing the coset representative.}. 
The matrix that defines the SO(4,4) group in our notation is
\be
\eta = \left(
\begin{array}{cc}
 \mathbf{0}_4 & \mathbf{1}_4  \\
  \mathbf{1}_4 & \mathbf{0}_4 \\
\end{array}
\right),
\ee
i.e., any $X \in$ SO(4,4) satisfies
\be
X^T \cdot \eta \cdot X = \eta.
\ee
The involution $\tilde \tau$ that defines\footnote{We put tilde of tau because it denotes an involution different from the standard Cartan involution.} the coset 
\be
\frac{\mathrm{SO}(4,4)}{\mathrm{SO}(2,2)\times \mathrm{SO}(2,2)}
\ee is:
\begin{align}
&\tilde\tau(\mathbb{H}_0) = -\mathbb{H}_0, 
& & \tilde\tau(\mathbb{H}_I) = - \mathbb{H}_I, \\
&\tilde\tau(\mathbb{E}_0) = + {\mathbb{F}_0}, 
& &\tilde\tau(\mathbb{E}_I) = + \mathbb{F}_I, \\
&\tilde\tau(\mathbb{E}_{q_{0}}) = + \mathbb{F}_{q_{0}}, 
& &\tilde\tau(\mathbb{E}_{q_{0}}) = - \mathbb{F}_{q_{I}}, \\
&\tilde\tau(\mathbb{E}_{p^{I}}) = - \mathbb{F}_{p^{0}}, 
& &\tilde\tau(\mathbb{E}_{p^{I}}) = + \mathbb{F}_{p^{I}}.
\end{align}
In our basis the matrix $\eta'$ that implements the involution as, 
\be
\tilde\tau(x) =: - x^\sharp =  - \eta' \cdot x^T \cdot  \eta', \qquad \mbox{for all} \quad x \in \mf{so}(4,4),
\ee is
\be
\eta' = \verb+diagonal+\{-1, 1, -1, 1, -1, 1, -1, 1\}.
\ee

Metric \eqref{cmapst} is obtained through the Maurer-Cartan one-form $\theta = d \cV \cdot \cV^{-1}$ as follows. Defining,
\be
P = \frac{1}{2} \left(\theta - \tilde \tau(\theta) \right),
\ee
one sees that 
\be
G_{ab} d \varphi^a \wedge \star d \varphi^b = \mathrm{Tr}(P \wedge \star P).
\ee

\section{Coset description of Bena-Warner solutions}
\label{sec:BW_coset}
The above dimensional reduction is well adapted to obtain coset description of Bena-Warner~\cite{Bena:2005va} solutions, since Bena-Warner solutions are naturally written in a fibre form amenable to such a reduction.
In this section we work out such a coset description in our notation. Some of this discussion is implicit in the literature \cite{Bossard:2011kz}; however, in order to work out the Geroch group description, we need  those details explicitly.

To set the notation we start with a quick summary of the Bena-Warner class of solutions, following the original notation~\cite{Bena:2005va}. In eleven dimensions,  metric takes the form,
\be
ds^2_{11} = ds^2_{5} + ds^2_{\mathrm{T}^6}, 
\ee
where 
\be
ds^2_5 = - (Z_1 Z_2 Z_3)^{-2/3}(dt + k)^2 + (Z_1 Z_2 Z_3)^{1/3} h_{mn} dx^m dx^n, 
\ee
and the metric on six-torus is
\be
ds^2_{\mathrm{T}^6} = \left(\frac{Z_2 Z_3}{Z_1^2}\right)^{1/3} (d\tilde x_1^2 + d \tilde x_2^2) + \left(\frac{Z_1 Z_3}{Z_2^2}\right)^{1/3} (d\tilde x_3^2 + d\tilde x_4^2)+\left(\frac{Z_1 Z_2}{Z_3^2}\right)^{1/3} (d\tilde x_5^2 + d\tilde x_6^2), 
\ee
The eleven dimensional three-form-field takes the form,
\be
A_{[3]} = A^1_{[1]} \wedge d\tilde x_1  \wedge d\tilde x_2  +A^2_{[1]} \wedge d\tilde x_3  \wedge d\tilde x_4 +A^3_{[1]} \wedge d\tilde x_5  \wedge d\tilde x_6.
\ee

In the above expressions $k$ is a one-form on the four-dimensional base space $ds^2_4 = h_{mn} dx^m dx^n$ and $A^I_{[1]}$ ($I=1,2,3$) are one-forms in the five-dimensional spacetime.
The BPS equations of motion determine all these fields in terms of eight harmonic functions 
\be
(V, M, K^I, L_I) \label{harmonic_functions}
\ee 
when the four-dimensional base metric $h_{mn}$ is taken to be the multi-center Gibbons-Hawking space
\cite{Gauntlett:2002nw, Gauntlett:2004qy, Bena:2005va, Berglund:2005vb}. We restrict our study to this case only. The relevant expressions are as follows. The four-dimensional base metric takes a fibre form,
\be
ds^2_4 = V^{-1}(d z + A)^2 + V (dr^2 + r^2 d\theta^2 + r^2 \sin^2 \theta d\phi^2),
\ee
with 
\be
\star_3 d A = dV, \label{GH_one_form}
\ee
and where the three-dimensional hodge star operation is with respect to the three-dimensional flat base metric 
\be
 ds^2_3 =  dr^2 + r^2 d\theta^2 + r^2 \sin^2 \theta d\phi^2.
 \ee
The five-dimensional vector fields take the form,
\be
A^I = - Z_I^{-1}(dt + k) + V^{-1}K^I (dz + A) + \xi^I,
\ee
with
\be
\star_3 d \xi^I = - d K^I, \label{duality_KI}
\ee
and the functions $Z_I$'s are given as
\be
Z_I = \frac{1}{2} C_{IJK} V^{-1} K^J K^K + L_I.
\ee
The one form $k$ on the four-dimensional Gibbons-Hawking space takes the form
\be
k = \mu (dz + A) + \omega_\rom{BW} \label{k_form},
\ee
with
\be
\mu = \frac{1}{6}C_{IJK} \frac{K^I K^J K^K}{V^2} + \frac{1}{2V} K^I L_I + M, \label{formula_mu}
\ee
and
\bea
\star_3 d \omega_\rom{BW} &=& V d \mu - \mu dV - V Z_I d \left( V^{-1} K^I\right) \label{omega_BW_1}
 \\ &=& V dM - M dV + \frac{1}{2}(K^I dL_I - L_I dK^I). \label{omega_BW_2}
\label{omega_BW}
\eea
To go from \eqref{omega_BW_1} to \eqref{omega_BW_2} we have substituted the formula \eqref{formula_mu} for the function $\mu$.  We use the notation $\omega_\rom{BW}$ to denote the three-dimensional one-form that appears in \eqref{k_form}, in order to distinguish it from the three-dimensional one-form  $\omega_3$ that we introduced in equation \eqref{an3} in section \ref{sec:dimred}. 

We refer the reader to \cite{Bena:2005va} and to reviews \cite{Bena:2007kg, Chowdhury:2010ct}  for further details on the solutions and the BPS equations. 

\subsection{SO(4,4) coset description}
\label{sec:current}

The above class of solutions can be viewed as solutions to five-dimensional U(1)$^3$ supergravity upon dimensional reduction from eleven to five dimensions on the six-torus. These solutions also have $\partial_t$ and $\partial_z$ Killing symmetries. As a result, they can be given a three-dimensional description in the dimensionally reduced U(1)$^3$ supergravity theory with three dimensional flat base space. Our first aim is to obtain that description, and in particular, how they are represented in the SO(4,4) coset variables. To this end we compute the \emph{sixteen} three-dimensional scalars that parameterise the coset
\be
\frac{\mathrm{SO}(4,4)}{\mathrm{SO}(2,2) \times \mathrm{SO}(2,2)},
\ee
in terms of the above introduced \emph{eight} harmonic functions \eqref{harmonic_functions}. 
Note that the sixteen three-dimensional scalars are
\be 
\{U, \sigma, \mathrm{Re} (X^I) =- \chi^I, \mathrm{Im} (X^I) =  f h^I, \zeta^\Lambda, \tilde \zeta_\Lambda \}.
\ee 
We find in the notation of section \ref{sec:dimred} for the scalars $\mathrm{Im} (X^I) = f h^I$,
\bea
h^1 &=& \left(\frac{Z_2 Z_3}{Z_1^2}\right)^{1/3}\\
h^2 &=& \left(\frac{Z_1 Z_3}{Z_2^2}\right)^{1/3} \\
h^3 &=& \left(\frac{Z_1 Z_2}{Z_3^2}\right)^{1/3}
\eea
together with 
\be
f =  (Z_1 Z_2 Z_3)^{-1/3},  
\ee
For the scalars  $\mathrm{Re} (X^I) =- \chi^I $ we find
\be
 \chi^I = - Z_I^{-1}.
\ee

With these fields at hand we can compute the  $\mbox{Re} N$ and $\mbox{Im} N$ matrices. We find the following simple expressions for these matrices
\be
\mbox{Re} N = \left(
\begin{array}{cccc}
 -\frac{2}{Z_1 Z_2 Z_3} & \frac{1}{Z_2 Z_3} &
   \frac{1}{Z_1 Z_3} & \frac{1}{Z_1 Z_2} \\
 \frac{1}{Z_2 Z_3} & 0 & -\frac{1}{Z_3} & -\frac{1}{Z_2} \\
 \frac{1}{Z_1 Z_3} & -\frac{1}{Z_3} & 0 & -\frac{1}{Z_1} \\
 \frac{1}{Z_1 Z_2} & -\frac{1}{Z_2} & -\frac{1}{Z_1} & 0 \\
\end{array}
\right),
\ee
and
\be
\mbox{Im} N =\left(
\begin{array}{cccc}
 -\frac{2}{Z_1 Z_2 Z_3} & \frac{1}{Z_2 Z_3} &
   \frac{1}{Z_1 Z_3} & \frac{1}{Z_1 Z_2} \\
 \frac{1}{Z_2 Z_3} & -\frac{Z_1}{Z_2 Z_3} & 0 & 0 \\
 \frac{1}{Z_1 Z_3} & 0 & -\frac{Z_2}{Z_1 Z_3} & 0 \\
 \frac{1}{Z_1 Z_2} & 0 & 0 & -\frac{Z_3}{Z_1 Z_2} \\
\end{array}
\right).
\ee

Proceeding further we have for the scalars $\zeta^\Lambda$
\bea
\zeta^0 &=& \mu, \\
\zeta^I &=& V^{-1} K^I, \label{scalar_zeta_I}
\eea
and for the scalar $U$,
\be
e^{2U}=V^{-1}.
\ee

Out of sixteen we have listed eleven scalars so far. Now we use the dualization equations \eqref{dual1}  and \eqref{dual2} to find the remaining five scalars $\tilde \zeta_\Lambda$ and $\sigma$. For evaluating $\tilde \zeta_\Lambda$  we need the combinations $\star_3 (F_3^\Sigma + \zeta^\Sigma F_3)$. Interestingly, these three-dimensional hodge dualities can be performed rather straightforwardly given the equations above. To this end, we first write the relation between the five three-dimensional one-forms $(\omega_3, A^\Lambda_{3})$ that appear in the coset description and the three-dimensional one forms that appear in Bena-Warner solutions $(A, \omega_\rom{BW}, \xi^I)$:
\bea
\omega_3 &=& A, \\
A_3^0 &= & \omega_\rom{BW}, \\
A_3^I &=&  \xi^I.
\eea
For $ \star_3 (F_3^0 + \zeta^0 F_3) $, we have
\bea
 \star_3 ( F_3^0 + \zeta^0 F_3) &=&  \star_3 (d\omega_\rom{BW}+ \zeta^0 dA) \\
& =& \star_3 d\omega_\rom{BW} + \zeta^0 dV  \\
&=& V d \mu - V Z_I d \zeta^I,
\eea
where in the first step we have used $\omega_3 = A$ and $A_3^0 = \omega_\rom{BW}$; in the second  step we have used the duality relation \eqref{GH_one_form}; and finally in the third step we have used relations 
\eqref{omega_BW_1}
and
\eqref{scalar_zeta_I}. Similarly, 
\bea
 \star_3 (F_3^I + \zeta^I F_3) &=& \star F_3^I + V^{-1}K^I dV \\ 
 &=& - d K^I + V^{-1}K^I dV,
\eea
where we have used $\omega_3 = A$ and $A_3^I = \xi^I$ together with relations
\eqref{scalar_zeta_I}  and  \eqref{duality_KI}.

 Using these expressions, an explicit calculation shows that the combination
 \be
 e^{2U}(\mathrm{Im}N)_{\Lambda \Sigma} \star_3 (F_3^\Sigma + \zeta^\Sigma F_3) - (\mathrm{Re}N)_{\Lambda \Sigma} d\zeta^\Sigma
 \ee
 vanishes. Therefore, the four $\tilde \zeta_\Lambda$ scalars can all be chosen to be zero, which is what we will do:
 \be
 \tilde \zeta_\Lambda =0.
 \ee
  Doing so we have from \eqref{dual2}
 \be
 d \sigma   = - 2 V^{-2} \star_3 F_3 = - 2 V^{-2} dV = 2 d (V^{-1}),
 \ee
 i.e., $\sigma$ can be taken to be $2 V^{-1}$:
 \be
 \sigma =2 V^{-1}.
 \ee
  To summarise, the sixteen scalars of the coset model take the values
 \bea
 e^{2U} &=& V^{-1}, \\
 \sigma &=& 2 V^{-1}, \\ 
 \tilde \zeta_\Lambda &=& 0, \\
\zeta^0 &=& \mu~~=~~\frac{1}{6}C_{IJK} \frac{K^I K^J K^K}{V^2} + \frac{1}{2V} K^I L_I + M, \\
\zeta^I &=& V^{-1} K^I, \\
\mathrm{Re} (X^I)  &=& - \chi^I~~=~~Z_I^{-1} = \left(\frac{1}{2} C_{IJK} V^{-1} K^J K^K + L_I\right)^{-1}, \\
\mathrm{Im} (X^I)  &=& f h^I~~=~~Z_I^{-1} = \left(\frac{1}{2} C_{IJK} V^{-1} K^J K^K + L_I\right)^{-1}.
 \eea

 Note that the right hand side of the above expressions are all written in terms of the eight Bena-Warner harmonic functions \eqref{harmonic_functions}. These expressions provide the required embedding of the Bena-Warner class of the solutions in SO(4,4) coset model framework in our notation. 
 The matrix of scalars 
 \be
 S(\vec x) = \cV^\sharp \cV
 \ee for the Bena Warner solutions takes the form,
\be
S(\vec x) = \left(
\begin{array}{cccccccc}
S_{11} & L_2 &S_{13} &
   K^1 & -1 & L_3 & 0 & -2 M \\
 -L_2 & 0 & K^3 & 0 & 0 & -1 & 0 & 0 \\
S_{13} & -K^3 &S_{33} & V
   & 0 & -K^2 & -1 & L_1 \\
 -K^1 & 0 & -V & 0 & 0 & 0 & 0 & -1 \\
 -1 & 0 & 0 & 0 & 0 & 0 & 0 & 0 \\
 -L_3 & -1 & K^2 & 0 & 0 & 0 & 0 & 0 \\
 0 & 0 & -1 & 0 & 0 & 0 & 0 & 0 \\
 2 M & 0 & -L_1 & -1 & 0 & 0 & 0 & 0 \\
\end{array}
\right), \label{matrix_scalars}
\ee
where
\bea
S_{11} &=& L_2 L_3 - 2 K^1 M, \\
S_{13} &=& \frac{1}{2} \left(K^1 L_1 - K^2 L_2 - K^3 L_3  - 2 M V \right), \\
S_{33} &=& K^2 K^3 + L_1 V.
\eea
We use $\vec x$ to collectively denote three-dimensional base space coordinates $r, \theta, \phi$. Using this embedding we can now arrive at some general results.
 
Our first general result is that the Lie algebra valued Noether's current for the Bena-Warner solutions is nilpotent. This is seen by a simple calculation as follows. We compute 
\be
J =  S^{-1} \cdot dS ,  \label{current_1}
\ee 
and find
\be
J =  dK^I \mathbb{F}_{p^{I}} -dV {\mathbb{F}_0} - dL_I \mathbb{F}_I - 2 dM \mathbb{E}_{q_{0}} + \left( M dV-  V d M  +\frac{1}{2}(L_I dK^I - K^I d L_I )\right) \mathbb{F}_{p^{0}}.  \label{current_2}
\ee
Clearly
\be
d \star_3 J = 0, \label{EOM}
\ee
since (i) functions \eqref{harmonic_functions} are harmonic, and (ii) from \eqref{omega_BW} we see that the term in front of $\mathbb{F}_{p^{0}}$ is $\star_3 d \omega_\rom{BW}$. Therefore, 
$J$ is a conserved Lie algebra valued one-form. It is easily checked that as an $\mathfrak{so}(4,4)$ matrix it is nilpotent of degree three. We define the charge matrix
\be
\mathcal{Q} = \frac{1}{4\pi}\int_\Sigma \star_3 J, \label{charge_matrix}
\ee
where $\Sigma$ is a two-cycle in three-dimensional flat space. If there are no Dirac-Misner strings in the solutions, then in the charge integral
there will not be a contribution from the term
\be
\mathbb{F}_{p^{0}} \int_\Sigma d \omega_\rom{BW}, \label{smooth}
\ee
as the integral is over a two cycle of a closed two form.

We can also connect  the preceding discussion  to that of reference \cite{Bossard:2011kz}. The four-dimensional (Lorentzian) gravity embedded in the SO$(4,4)$ coset is  parameterised by $f$ and 
$\tilde \zeta^0$, while other scalars are to be set to zero. For this embedding, we have the gravity $\mathfrak{sl}(2)$ spanned by the generators
\bea
\mathbf{h}&=& -\frac{1}{2} \left(\mathbb{H}_0 + \sum_{I=1}^{3} \mathbb{H}_I\right), \label{sl2H} \\
\mathbf{e} &=& \mathbb{F}_{p^{0}}, \\
\mathbf{f} &=& \mathbb{E}_{p^{0}}.
\eea
The $\mathbf{h}$ gives the following grading for the generators that appear in current \eqref{current_2}:
\be
(\mathbb{F}_{p^{0}})^{(+2)}, \{ {\mathbb{F}_0}, \mathbb{F}_I, \mathbb{E}_{q_0}, \mathbb{F}_{p^I} \}^{(+1)}. \label{grading}
\ee
For the equations of motion \eqref{EOM}, we observe that coefficients at grade one are precisely the equations giving eight harmonic functions; moreover, coefficient of $J$ at grade two is related to the no Dirac-Misner strings condition \cite{Bossard:2011kz}.

\subsection{Gauge transformations and spectral flow dualities}

The Bena-Warner class of solutions have two well studied symmetries called ``gauge transformations'' and ``spectral flows''. Under these  transformations solutions are mapped to solutions. The first class, the so called ``gauge transformations''  correspond to the changing the harmonic  functions as \cite{Bena:2005va}
\begin{align}
& V & &\to & & V, &  \\
&K^I & &\to & & K^I + c^I V, & \\
& L_I & &\to & & L_I - C_{IJK} c^J K^K - \frac{1}{2} C_{IJK} c^J c^K V, & \\
 &M & & \to &  & M -\frac{1}{2} c^I L_I + \frac{1}{12} C_{IJK} ( V c^I c^J c^K + 3 c^I c^J K^K).&
\end{align}
In the notation of coset variables, these transformations correspond to shifting the scalars $\zeta^I$ by constant $c^I$. Indeed, these gauge transformations are realised as simply
\be
S(\vec x) \to g^\sharp S(\vec x) g, 
\ee
with
\be
g = \exp \left[ - \sum_{I=1}^{3} c^I \mathbb{E}_{{q_I}}\right]. \label{shift}
\ee

The second class of transformations are called the spectral flow transformations \cite{Bena:2008wt}. They are
\begin{align}
& M & & \to& & M,& \\
& L_I & &\to&& L_I - 2 c_I M, & \\
& K^I & &\to&& K^I - C^{IJK} c_J L_K  +  C_{IJK} c_J c_K M, &\\
& V & &\to&&  V + c^I K_I - \frac{1}{2}C^{IJK}c_I c_J L_K + \frac{1}{3} C^{IJK} c_I c_J c_K M. &
\end{align}
These transformations do not correspond a simple shifting of scalars, instead these are Harrison transformations -- transformations involving ``negative'' root vectors of the $\mathfrak{so}(4,4)$ Lie algebra,
\be
S(x) \to g^\sharp S(x) g, 
\ee
with
\be
g = \exp \left[ - \sum_{I=1}^{3} c^I \mathbb{F}_{{q_I}}\right]. \label{harrison}
\ee
Harrison transformations have non-trivial action on the solutions \cite{Virmani:2012kw, Sahay:2013xda}. One can clearly come up with a very large class of similar other transformations by exponentiating other generators.\footnote{The Lie algebra generators appearing in equations \eqref{shift} and \eqref{harrison} are all at grade zero with respect to generator $\mathbf{h}$ defined in \eqref{sl2H}, i.e., $\mathbb{E}_{{q_I}}$ and $\mathbb{F}_{{q_I}}$ also belong to the Lie algebra of the four-dimensional U-duality group.}

With this  understanding of spectral flows as Harrison transformations, we are also able to better understand relation between references \cite{Bena:2012wc, Cvetic:2013vqi} and \cite{Virmani:2012kw, Sahay:2013xda}.  On the one hand, reference~\cite{Cvetic:2013vqi} uses the so-called  TsT transformations and some generalisations to relate certain asymptotically flat black hole solutions to their subtracted geometries. It was previously understood that those TsT transformations are related to spectral flows \cite{Bena:2012wc}. On the other hand,  references~\cite{Virmani:2012kw, Sahay:2013xda} use a class of Harrison transformations to relate the same asymptotically flat black hole solutions to their subtracted geometries. Via  these inter-connections,  it was expected that Harrison transformations and closely related to spectral flows. Above, we have presented a direct and simple relationship: Bena-Bobev-Warner spectral flows are a special class of Harrison transformations. To the best of our knowledge, such a direct relationship  has not been elucidated before.\footnote{We thank Monica Guica for discussion on this point.} 

\section{Geroch group description of collinear bubbling solutions}
\label{sec:Geroch} 
Now we are in position to work out the Geroch group description of collinear bubbling solutions. We need to work with collinear centers, since only in this situation we can perform further dimensional reduction to two dimensions. We keep the initial discussion general, i.e., with finite set of isolated centers with location anywhere in  flat base $\mathbb{R}^3$. Later when we discuss reduction to two dimensions, we will assume $\partial_\phi$ as another Killing symmetry of the full spacetime configuration.

\subsection{Matrix of scalars for bubbling solutions}

In the Bena-Warner class of solutions, we can choose the harmonic functions to be localised anywhere on the three-dimensional base space. We take harmonic functions with finite set of isolated sources at locations $\vec x^{(j)}$ in the three-dimensional base space $\mathbb{R}^3$. For convenience of notation, in the equations below we put all $I, J, K$ indices upstairs, and the indices for the centers downstairs. We have for the eight harmonic functions,
\begin{align}
V &= q_0 + \sum_{j=1}^{N} \frac{q_j}{r_j}, & 
K^I &= k^I_0 +  \sum_{j=1}^{N} \frac{k^I_j}{r_j}, & \label{harm_defn1} \\
L^I &= l^I_0 +  \sum_{j=1}^{N} \frac{l^I_j}{r_j},& 
M &= m_0 +  \sum_{j=1}^{N} \frac{m_j}{r_j},\label{harm_defn2}
\end{align}
where $r_j = | \vec x - \vec x^{(j)}|$.
We can easily read off the various matrix elements of the matrix \eqref{matrix_scalars}. There are only three matrix elements that are slightly non-trivial. We calculate them here,
\bea
S_{11} &=&L_2 L_3 - 2 K^1 M \\
&=& \left(l^2_0 +  \sum_{s=1}^{N} \frac{l^2_s}{r_s}\right) \left(
l^3_0 +  \sum_{t=1}^{N} \frac{l^3_t}{r_t} \right) - 2 \left(k^1_0 +  \sum_{s=1}^{N} \frac{k_s^1}{r_s} \right) \left( m_0 +  \sum_{t=1}^{N} \frac{m_t}{r_t} \right)
\eea
Expanding it out we get,
\be
S_{11}= \left(l^2_0 l^3_0 - 2 k^1_0 m_0 \right) + \sum_{s=1}^{N} \frac{1}{r_s} \left(l^2_s l^3_0 +  l^2_0 l^3_s - 2 k^1_s m_0  - 2 k^1_0 m_s   \right)
+ \sum_{\substack{s,t={1}}}^{N} \frac{1}{r_s r_t} \left(l^2_s  l^3_t - 2 k^1_s m_t \right). \label{expand_S11}
\ee
In this expansion there \emph{are} double pole terms.

For bubbling solutions, i.e., solutions without brane sources, all harmonic functions must have singularities only at the Gibbons-Hawking centres. Such solutions have \cite{Bena:2005va}:
 \begin{align}
& l_j^I  =- \frac{1}{2} C_{IJK} \frac{k^J_j k^K_j}{q_j},  & 
& m_j  =\frac{1}{2} \frac{k^1_j k^2_j k^3_j}{q_j^2}. & \label{bubbling}
\end{align}
Since $q_j$ appears in the denominator in these equations, in order for these equations to make sense, we must have Gibbons-Hawking center at location $\vec x^{(j)}$, i.e., $q_j\neq 0$. For this sub-class  of solutions double pole terms in \eqref{expand_S11}  all cancel out:
\be
l^2_s  l^3_s - 2 k^1_s m_s = \left( - \frac{k^s_1 k^s_3}{q_s} \right)\left( - \frac{k^s_1 k^s_2}{q_s} \right) - (2 k^1_s) \left( \frac{1}{2} \frac{k^1_s k^2_s k^3_s}{q_s^2} \right) = 0.
\ee
Similarly, double pole terms all cancel out for $S_{13}$ and $S_{33}$. The expansion of $S_{13}$ is as follows,
\bea
S_{13} &=& \frac{1}{2} \left(K^1 L_1 - K^2 L_2 - K^3 L_3  - 2 M V  \right) \nn \\
&=&\frac{1}{2} \left(k^1_0 l^1_0 - k^2_0 l^2_0 - k^3_0 l^3_0  - 2 m_0 q_0  \right)  \nn \\ 
& &+  \sum_{s=1}^{N}  \frac{1}{2 r_s} \left(k^1_s l^1_0 - k^2_s l^2_0 - k^3_s l^3_0  - 2 m_s q_0 + k^1_0 l^1_s - k^2_0 l^2_s - k^3_0 l^3_s  - 2 m_0 q_s  \right) \nn \\
& &  +  \sum_{\substack{s,t=1, \\ s\neq t}}^{N} \frac{1}{2 r_s r_t} \left(k^1_s l^1_t - k^2_s l^2_t - k^3_s l^3_t  - 2 m_s q_t  \right),
\eea
and expansion of $S_{33}$ is as follows,
\bea
S_{33} &=& K^2 K^3 + L_1 V \nn \\ 
&=& \left(k^2_0 k^3_0 +  l^1_0 q_0 \right) + \sum_{s=1}^{N}  \frac{1}{r_s}  \left( k^2_s k^3_0 + k^2_0 k^3_s +  l^1_s q_0 +  l^1_0 q_s \right) +  \sum_{\substack{s,t=1, \\ s\neq t}}^{N} \frac{1}{r_s r_t} \left(k^2_s  k^3_t +  l^1_s q_t \right).
\eea
With these expansions at hand we can write the expanded $S(\vec x)$ matrix schematically as, 
\be
S(\vec x) = S_\infty+ \sum_{s=1}^{N} \frac{1}{r_s} S_s(\vec x) + \sum_{\substack{s,t=1, \\ s \neq t}}^{N} \frac{1}{r_s r_t} S_{st}. \label{matrix_S}
\ee
Properties of $S_s$ and $S_{st}$ matrices  are of interest to us. Remarkably, these matrices satisfy a number of simple properties. The key properties are,
\begin{enumerate}
\item The matrices $S_\infty^{-1} S_s$ are nilpotent of degree three. When  conditions \eqref{bubbling} for the bubbling solutions are imposed in addition, $S_\infty^{-1} S_s$ are nilpotent of degree two. Unless otherwise stated we will implicitly assume conditions \eqref{bubbling}. We define
\be
B_s := S_\infty^{-1} S_s, \qquad \qquad B_s \cdot B_s = 0.
\ee
\item Nilpotent matrices $B_s$ are in the $8\times8$ vector representation of the Lie algebra $\mf{so}(4,4)$. These matrices are of rank 2. In fact, these matrices as Lie algebra generators belong to the unique minimal nilpotent orbit of $\mf{so}(4,4)$. In the notation of the appendix \ref{app:nilpotent_orbits}, the orbit is $\mathcal{O}_1$ with $\alpha$ and $\beta$ labels (0,1,0,0) and (1,1,1,1) respectively. 
\item 
When we multiply three such matrices we get zero, 
\be
B_r \cdot B_s \cdot B_t = 0. 
\ee
\item The matrices $S_{st}$ are,
\be
S_{st} = \frac{1}{2}S_\infty \cdot (B_s \cdot B_t + B_t \cdot B_s).
\ee
\item The full matrix  $S(\vec x)$ is therefore simply
\be
S(\vec x) = S_\infty  \cdot \exp \left[ \sum_{s=1}^{N}\frac{B_s}{r_s}\right]. \label{S_exp}
\ee
\end{enumerate}
Some of these comments admit appropriate generalisation when the bubbling conditions \eqref{bubbling} are not imposed. However, since for such solutions matrices $B_s$ are not of rank 2, we do not expect them to fit into the inverse scattering approach of \cite{KKV2}. For this reason we have not explored that case in detail. We leave a more detailed study of those solutions for the future.

\subsection{Charge matrices at each center}

One can write  expressions for  $B_s$ matrices for any given set of harmonic functions of the form of (\ref{harm_defn1})--(\ref{harm_defn2}) by reading off the coefficients of $r_s^{-1}$ in $S(\vec x)$. We find
\be
B_s = - q_s \mathbb{F}_0 - l_s^I \mathbb{F}_I - \b_s \mathbb{F}_{p^0} + k_s^I \mathbb{F}_{p^I} - 2m_s \mathbb{E}_{q_0},          \label{formula_bs}
\ee 
where 
\be
\b_s = q_0m_j - m_0q_j + \frac12 (l^I_jk_0^I - k^I_jl^I_0).
\ee

To calculate the charge matrix at each center, we evaluate $J=S^{-1}dS$, and integrate on a small sphere around the $s^{\text{th}}$ center to find
\begin{align}
\cQ_{(s)} =&\ \frac1{4\pi}\int_{\Sigma_s}\star_3J = -B_s -  \frac{1}{2} \sum_{\substack{t=1 \\ t\neq s}}^N \frac{[B_s, B_t]}{|r_s-r_t|}.         \label{chargenew}
\end{align}
The commutator involved can be written in a nice way as well. One finds,
\be     \label{comm_bpbs}
[B_s,B_t] = q_s q_t \left(\frac{k^1_t}{q_t} - \frac{k^1_s}{q_s} \right)\left(\frac{k^2_t}{q_t} - \frac{k^2_s}{q_s} \right)\left(\frac{k^3_t}{q_t} - \frac{k^3_s}{q_s} \right) \  \mathbb{F}_{p^0}.
\ee

We have not used all regularity conditions up to now. The bubble equations require that \cite{Bena:2005va} 
\be
\frac{1}{2} \sum_{\substack{t=1 \\ t\neq s}}^N q_s q_t \left(\frac{k^1_t}{q_t} - \frac{k^1_s}{q_s} \right)\left(\frac{k^2_t}{q_t} - \frac{k^2_s}{q_s} \right)\left(\frac{k^3_t}{q_t} - \frac{k^3_s}{q_s} \right)\frac{1}{|r_{s}-t_{t}|} =  \b_s,
\ee 
which, using \eqref{comm_bpbs} becomes
\be
\frac{1}{2} \sum_{\substack{t=1 \\ t\neq s}}^N \frac{[B_s, B_t]}{|r_s-r_t|} = \b_s \mathbb{F}_{p^0}.
\ee
As a result, the charge matrix at each center of the bubbling solutions is then given by 
\be
\cQ_{(s)} =  q_s\mathbb{F}_0 + l_s^I\mathbb{F}_I - k_s^I \mathbb{F}_{p^I} + 2m_s \mathbb{E}_{q_0}.  \label{chargebubb}
\ee
It only depends on the charges at the location $\vec x^{(s)}$ and the Dirac-Misner strings have all cancelled upon using the bubble equations, as expected. Note that $\cQ_{(s)}$ is a linear combination of grade +1 terms alone, cf.~\eqref{grading}.

\subsection{Monodromy matrix for bubbling solutions}
\label{sec:bubble}
Let us now understand the above class of solutions from the Geroch group perspective. 
In order to reduce the solution to two dimensions, we need to take collinear centers; the matrix $S(\vec x)$ then only depends on two coordinates $(r, \theta)$.  A recipe for obtaining the Geroch group matrix $\cM(w)$ for a solution with known $S(\vec x)$ was given in  \cite{BM, Chakrabarty:2014ora}; see also \cite{Camara:2017hez}.
To use this recipe we first need to change coordinates of the base space from $(r, \theta, \phi)$ to the Weyl canonical coordinates $(\rho, z, \phi)$. The canonical coordinates have the property that the three-dimensional base space has the cylindrical metric
\be
ds_3^2 = d\rho^2 + d z^2 + \rho^2 d \phi^2.
\ee
The change of coordinates that takes from flat base space in polar coordinates $ds_3^2 = dr^2 + r^2 d \theta^2 + r^2 \sin^2 \theta d \phi^2$
to canonical coordinates is:
\begin{align}
\rho &= r \sin \theta, &
z &= r \cos \theta.
\end{align}
In canonical coordinates recipe to obtain Geroch group matrix is:
\be
\cM(w) = S(\rho =0,  \ z = w \quad \mbox{with} \quad z < -R), \label{curlyM_str_M}
\ee 
where $R$ is the radius of the  sufficiently large semicircle in the $(\rho, z)$ half-plane, such that all ``corners'' that feature in the ``rod-structure'' of a given solution are inside this semicircle. We refer the reader to \cite{Chakrabarty:2014ora} for a more precise discussion of these phrases.  For practical calculations we simply take the limit $\rho = 0$, $z$ near $-\infty$. The ``rod-structure'' of the Bena-Warner solutions has been recently studied in detail by Breunh\"older and Lucietti \cite{Breunholder:2017ubu}.

To use the recipe, we need to understand how various functions of $(r, \theta)$ appearing in $S(\vec x)$ change into functions of $w$. 
Let the centres be located along the third axis
\be
\vec x^{(j)} = \vec R_j = (0,0,w_j).
\ee
With this convention the functions $r_j$ become,
\be
r_j = |\vec x - \vec x^{(j)}| = \sqrt{r^2 - 2 r w_j \cos \theta + w_j^2}.
\ee
In the Weyl canonical coordinates 
\be
r_j =\sqrt{\rho^2 + (z-w_j)^2}.
\ee As a result, the harmonic functions $\frac{1}{r_j}$ upon taking the limit $\rho = 0$, $z$ near $-\infty$ takes the form
\be
\frac{1}{r_j} \longrightarrow - \frac{1}{w - w_j}. \label{replacement}
\ee
Using replacement \eqref{replacement} we get the matrix $\cM(w)$ from matrix \eqref{matrix_S}, 
\be
\cM(w) = S_\infty - \sum_{s=1}^{N} \frac{S_s(\vec x)}{w-w_s}  + \sum_{\substack{s,t=1 \\ s \neq t}}^{N} \frac{S_{st}}{(w-w_s)(w-w_t)}.
\ee
A more useful expression can be obtained using \eqref{S_exp}, and expanding the exponential.
\be
\cM(w) =  S_\infty \cdot \left( \mathbb{I}- \sum_{s=1}^{N} \frac{B_s}{w-w_s} + \frac{1}{2} \sum_{\substack{s,t=1 \\ s\neq t}}^{N} \frac{ \{B_s, B_t\}}{(w-w_s)(w-w_t)} \right),
\ee
We have a final expression 
\be
S_\infty^{-1} \cdot \cM(w) = \mathbb{I} + \sum_{s=1}^{N} \frac{A_s}{w-w_s},              \label{curlyM}
\ee
where
\be
A_s = - B_s +\frac12 \sum_{\substack{t=1 \\ t\neq s}}^{N} \frac{1}{w_s - w_t}   \{B_s, B_t\}. \label{residues}
\ee
This is one of our main results. We have shown that the Geroch group matrices \eqref{curlyM} for bubbling solutions only have simple poles. 
In the next subsection we explore further properties of the residue matrices.

\subsection{Properties of the monodromy matrix for bubbling solutions}

We start by recalling an elementary result from linear algebra that rank of an $n \times n$ matrices $A$ is equal to \be \mathrm{rank} (A) =  n - \mathrm{dim~of~null~space~of~}A.\ee 
Using this we argue that rank of $A_s$ is same as rank of $B_s$. Consider an arbitrary vector $v$ in the null space of $B_s$: 
$B_s v = 0.$
It follows that 
\be
w = \left(\mathbb{I} + \frac12 \sum_{\substack{t=1 \\ t\neq s}}^{N} \frac{1}{w_s - w_t}   B_t \right) \cdot v, \label{map_v_w}
\ee
is in the null space of $A_s$. To see this simply consider,
\bea
A_s w &=& A_s v +   \frac12 \sum_{\substack{t=1 \\ t\neq s}}^{N} \frac{1}{w_s - w_t}   A_s B_t v  \\
& = & \frac12 \sum_{\substack{t=1 \\ t\neq s}}^{N} \frac{1}{w_s - w_t}  B_s B_t v +   \frac12 \sum_{\substack{t=1 \\ t\neq s}}^{N} \frac{1}{w_s - w_t}   A_s B_t v \\
& = & \frac12 \sum_{\substack{t=1 \\ t\neq s}}^{N} \frac{1}{w_s - w_t}  B_s B_t v -   \frac12 \sum_{\substack{t=1 \\ t\neq s}}^{N} \frac{1}{w_s - w_t}   B_s B_t v\\
&=&0
\eea
where we have used $B_s \cdot B_t \cdot B_p = 0$. The map \eqref{map_v_w} is maximal rank (invertible):
\be
 \left(\mathbb{I} + \frac12 \sum_{\substack{t=1 \\ t\neq s}}^{N} \frac{B_t}{w_s - w_t}    \right)^{-1} = \left(\mathbb{I} - \frac12 \sum_{\substack{t=1 \\ t\neq s}}^{N} \frac{B_t}{w_s - w_t}    + \frac{1}{8}\sum_{\substack{t,r=1 \\ t,r\neq s}}^{N} \frac{ \{B_t, B_r\}}{(w_s - w_t)(w_s - w_r)}  \right), \label{inv_map}
\ee 
so every $v$ there is a $w$.

For the converse, consider a vector $ w$ in the null space of $A_s$. Then it follows that we can find a vector $v$, using  \eqref{inv_map}, that belongs to the null space of $B_s$. Consider:
\begin{equation}
0=A_s w =  -B_s w + \frac12 \sum_{t \neq s}^N \frac{\{B_s,B_t\}}{w_s-w_t} w.   \label{nullvec}      
\end{equation}
This is the same as 
\begin{equation}
B_s  w = \frac12 B_s \sum_{t \neq s}^N \frac{B_t}{w_s-w_t} w + \frac12 \sum_{t \neq s}^N\frac{B_t}{w_s-w_t} B_s w.
\end{equation}
Multiplying both sides by $B_p$ for any $p$ makes the RHS vanish. So we get that 
\begin{equation}
B_p B_s  w = 0  \qquad \forall\, w  \qquad  \text{s.t.}  \qquad  A_s  w =0,
\end{equation}
which in turn implies
\bea
0 = A_s w &=&  - B_s \cdot  \left(\mathbb{I} - \frac12 \sum_{\substack{t=1 \\ t\neq s}}^{N} \frac{B_t}{w_s - w_t}   \right) w\\
&=& - B_s \cdot  \left(\mathbb{I} - \frac12 \sum_{\substack{t=1 \\ t\neq s}}^{N} \frac{B_t}{w_s - w_t}    + \frac{1}{8}\sum_{\substack{t,r=1 \\ t,r\neq s}}^{N} \frac{ \{B_t, B_r\}}{(w_s - w_t)(w_s - w_r)}  \right) w .
\eea

It follows that generically $A_s$ is also a rank-2 matrix. This is our second main result. We have shown that the Geroch group matrices \eqref{curlyM} for bubbling solutions only have simple poles with residues of rank-2. It is precisely for this setup that the Riemann-Hilbert factorisation was developed in reference \cite{KKV2}. We leave the explicit factorisation of these matrices for future work.  Due to the property that $B_s \cdot B_p \cdot B_q = 0$, it follows that $A_s$ matrices are also nilpotent of degree 2. Generically, the $A_s$ matrices  do not belong to the Lie algebra $\mathfrak{so}(4,4)$, due to the presence of the anti-commutator terms in \eqref{residues}.

\section{Some explicit examples}
\label{sec:examples}
We now discuss some simple examples of Bena-Warner solutions and obtain their Geroch group matrices. 
\subsection{Supertube in Taub-NUT and the related bubbling geometry}
\subsubsection*{Supertube in Taub-NUT}
Our first example is one supertube in Taub-NUT \cite{Bena:2005ay}. 
The eight harmonic functions take the form in the notation of \cite{Bena:2008wt}
\begin{align}
&K^1   = 0, &
&K^2 = 0, & 
&K^3 = - \frac{q_3}{2\Sigma}  & 
&L_1  = 1 + \frac{Q_1}{4\Sigma},  & \\
&L_2   = 1 + \frac{Q_2}{4\Sigma}, &
&L_3  =    1 & 
&V  =  \epsilon_0 + \frac{1}{r}, &
&M  = \frac{J_T}{16R} -  \frac{J_T}{16}\frac{1}{\Sigma} &
\end{align}
where $\vec R = (0,0,R)$ is the position of the round supertube in Taub-NUT along the Taub-NUT fibre, and 
\be
\Sigma = | \vec r - \vec R| =  \sqrt{r^2 - 2rR \cos \theta + R^2}.
\ee
The requirement of the smoothness of the solution results in \cite{Bena:2005ay, Bena:2008wt}
\bea
q_3 J_T &=& Q_1 Q_2,  \label{smoothness_two_center_1}\\
J_T \left( \epsilon_0 + \frac{1}{R}\right)&=& 4q_3.  \label{smoothness_two_center_2}
\eea
This last condition is to be thought of as a condition on the separation $R$ in terms of charges. 

Charges for the configuration at the centre $\vec r=0$ and the centre $\vec r=\vec R$ are
\bea
\{q_1, l_1^I, k_1^I, m_1\}  &=&  \{1,0,0,0,0,0,0,0 \},\\
 \{q_2, l_2^I, k_2^I, m_2\} &=&  \{0,\frac{1}{4}Q_1,\frac{1}{4}Q_2,0,0,0,-\frac{1}{2}q_3, -\frac{1}{16}J_T \}.
\eea
We note that the Gibbons-Hawking charge at the second centre is zero $(q_2 = 0)$, though  M2 charges $l_1, l_2$ are non-zero. Thus, there are ``brane sources'' in this solution, it is not a bubbling solution. As a consequence, the general analysis of bubbling solutions given in section \ref{sec:bubble} does not apply. Nevertheless, the solution admits a very similar Geroch group description which we elucidate.

The charge matrices at the two centres can be calculated  using the analysis of section \ref{sec:current}, cf. \eqref{charge_matrix}. The general form of the charge matrix \eqref{charge_matrix} when the two-cycle is taken to be a two-sphere enclosing only $j$th center is: 
\be
\mathcal{Q}_{(j)} = -q_j \mathbb{F}_0 - l_j^I \mathbb{F}_I + k_j^I \mathbb{F}_{p^{I}} - 2 m_j \mathbb{E}_{q_{0}} \label{charge_matrix_2}, 
\ee
where we have used the smoothness condition;  cf.~discussion around equation \eqref{smooth}.
Therefore for the above example we have
\bea
\cQ_{(1)} &=& -\mathbb{F}_0, \\
\cQ_{(2)} &=&   - \frac{1}{4} Q_1 \mathbb{F}_1 -\frac{1}{4} Q_2 \mathbb{F}_2  -\frac{q_3}{2} \mathbb{F}_{p_{3}}  +  \frac{J_T}{8} \mathbb{E}_{q_{0}},
\eea

To obtain the Geroch group matrix for the above solution, we need to first find the matrix of scalars $S(\vec x)$ and then relate $S(\vec x)$ with $\cM(w)$ via \eqref{curlyM_str_M}. Writing as above
\be
S(\vec{x}) = S_\infty\cdot \exp\left[\frac{B_1}{r} + \frac{B_2}{|\vec{r}-\vec{R}|} \right],
\ee
and using (\ref{formula_bs}), we have
\begin{align}
B_1 =&  \ -\mathbb{F}_0 + \frac{J_T}{16R}\mathbb{F}_{p^0}     ,                   \\
B_2 =&\ -\frac{Q_1}{4}\mathbb{F}_1 - \frac{Q_2}{4}\mathbb{F}_2  - \frac{q_3}{2}\mathbb{F}_{p^3} + \frac{J_T}{8} \mathbb{F}_{q_0}- \frac{J_T}{16R} \mathbb{F}_{p^0},
\end{align}
and
\be
S_\infty =\left(
\begin{array}{cccccccc}
 1 & 1 & -\frac{J_T \epsilon_0}{16 R} & 0 & -1 & 1 & 0 &
   -\frac{J_T}{8 R} \\
 -1 & 0 & 0 & 0 & 0 & -1 & 0 & 0 \\
 -\frac{J_T \epsilon_0}{16 R} & 0 & \epsilon_0 & 
 \epsilon_0 & 0 & 0 & -1 & 1 \\
 0 & 0 & -\epsilon_0 & 0 & 0 & 0 & 0 & -1 \\
 -1 & 0 & 0 & 0 & 0 & 0 & 0 & 0 \\
 -1 & -1 & 0 & 0 & 0 & 0 & 0 & 0 \\
 0 & 0 & -1 & 0 & 0 & 0 & 0 & 0 \\
 \frac{J_T}{8 R} & 0 & -1 & -1 & 0 & 0 & 0 & 0 \\
\end{array}
\right).
\ee
From these expression it is easy to verify equation (\ref{chargenew}) 
\begin{align}
\cQ_{(1)}   &= - B_1 - \frac{1}{2R} [B_1,B_2],           &
\cQ_{(2)}   &= - B_2 + \frac{1}{2R} [B_1,B_2].
\end{align}
Now, from relations (\ref{residues}) one can now readily calculate matrices $A_1,\,A_2$:
\begin{align}
A_{(1)}   &= - B_1 - \frac{1}{2R} \{B_1,B_2\},           &
A_{(2)}   &= - B_2 + \frac{1}{2R} \{B_1,B_2\},
\end{align}
and the monodromy matrix is 
\be
\cM(w) = S_\infty \cdot \left(\mathbb{I} +   \frac{A_1}{w} +   \frac{A_2}{w-R}\right).
\ee
The $A_1,\,A_2$ matrices  satisfy several properties required for inverse scattering construction \cite{KKV2}  to work, e.g.,
\begin{align}       \label{cond_inv_scatt}
A_k^\sharp &= S_\infty\cdot A_k\cdot S_\infty^{-1},&
A_k \cdot \eta  \cdot A_k^T &=  0.
\end{align}

\subsection*{Spectral flowed bubbling geometries}
A bubbling solution associated to the above configuration  is obtained by applying a combination of gauge and spectral flow transformations \cite{Bena:2008wt}. The harmonic functions are given as   
\begin{align}
K_1&=\ -\gamma  - \frac{\gamma  Q_2}{4\S}, & 
K_2 &= -\gamma  - \frac{\gamma  Q_1}{4\S},& \\
K_3 &= \left(\e_0+\frac1r\right)\frac{q_3}2 - \frac{q_3}{2\S},  &  
L_1& = 1+\frac{Q_1}{4\S},   & \\
L_2& = 1+\frac{Q_2}{4\S},& 
L_3 &= 1+\frac{\gamma q_3}2 - \frac{\gamma J_T}{8R} +  \frac{\gamma J_T}{8\S},    &    \\
V &= \left(\e_0+\frac1r\right)\left( 1 + \frac{\gamma q_3}2 \right) - \frac{\gamma q_3}{2\S},      &
M &=  \frac{J_T}{16R} - \frac{q_3}4 - \frac{J_T}{16\S}.        &
\end{align}
The charges at the centers are 
\begin{align}
 \{q_1, l_1^I, k_1^I, m_1\} &=  \left\{ 1+\frac{\gamma q_3}2,0,0,0,0,0,\frac{q_3}2,0   \right\},      \\
  \{q_2, l_2^I, k_2^I, m_2\}& =  \left\{ -\frac{\gamma q_3}2,\frac{Q_1}4,\frac{Q_2}4, \frac{\gamma J_T}8, -\frac{\gamma Q_2}4, 
-\frac{\gamma Q_1}4,-\frac{q_3}2, -\frac{J_T}{16}   \right\},
\end{align}
and we still have the smoothness conditions (resulting from the bubble equations now) (\ref{smoothness_two_center_1}) and (\ref{smoothness_two_center_2}). Again, we write the matrix of scalars as an exponential of matrices $B_1, B_2$ as before, finding
\bea
B_1 &=& -\left(1 + \frac{\gamma q_3}{2} \right) \mathbb{F}_0 + \frac{J_T}{16R} \mathbb{F}_{p^0} + \frac{q_3}{2} \mathbb{F}_{p^3}, \\
B_2 &=& \frac{\gamma q_3}{2} \mathbb{F}_0 - \frac{Q_1}{4} \mathbb{F}_1 - \frac{Q_2}{4} \mathbb{F}_2 - \frac{\gamma J_T}{8}\mathbb{F}_3   - \frac{J_T}{16R}\mathbb{F}_{p^0} - \frac{\gamma Q_2}{4}\mathbb{F}_{p^1}  \nn \\ 
&&   - \frac{\gamma Q_1}{4}\mathbb{F}_{p^2} - \frac{q_3}{2} \mathbb{F}_{p^3}  + \frac{J_T}{8}\mathbb{E}_{q_0}.
\eea
Matrices $\cQ_{(1)}, \cQ_{(2)}$ and $A_1, A_2$ and can now be readily computed. 
Once again matrices $A_1$ and $A_2$ matrices satisfy properties required for inverse scattering construction of \cite{KKV2} to work.

\subsection{Many supertubes in Taub-NUT and the related bubbling geometries}

The above example is easily generalised to $N$ two-charge supertubes in Taub-NUT.  Such a configuration is specified by the following eight harmonic functions \cite{Bena:2005ay}
\begin{align}
V & =  \epsilon_0 + \frac{1}{r}, &
K^1  & = 0, &
K^2  &= 0, &
K^3 &= k_0^3- \sum_{i=1}^{N}\frac{q_3^i}{2r_i} & \\
M  &= m_0- \sum_{i=1}^{N}\frac{J_i}{16r_i},&
L_1  &= l_0^1 + \sum_{i=1}^{N}\frac{Q_1^i}{4r_i}, &
L_2   &= l_0^2 + \sum_{i=1}^{N}\frac{Q_2^i}{4r_i}, &
L_3  &=    l_0^3,&
\end{align}
where $r_i=|\vec{r}-\vec{R}_i|$, with $\vec{R}_i = (0,0,R_i)$ the positions of the supertubes. Smoothness conditions give $N$ relations \cite{Bena:2005ay}, 
\be
Q_1^iQ_2^i=q_3^iJ_i, \label{smoothness_eqs}
\ee
and the condition that the solution be free of Dirac-Misner strings at each center gives the following $N+1$ conditions, for $i=1,\ldots, N,$
\be
\left(\e_0+\frac{1}{R_i}\right)J_i = 4l_0^3q_3^i,   \qquad \quad m_0=\frac{1}{16}\sum_{i=1}^{N}\frac{J_i}{R_i}. \label{bubbling_eqs}
\ee
For this configuration,  matrices $B_j$ for $j=1,\ldots, N,$ are
\be
B_j = -\frac{Q_1^j}4 \mathbb{F}_1 - \frac{Q_2^j}4 \mathbb{F}_2  - \frac{q_3^j}2 \mathbb{F}_{p^3} + \frac{J_j}8 \mathbb{E}_{q^0} - \frac{J_j}{16R_j} \mathbb{F}_{p^0},
\ee
and the matrix $B_0$ for the Taub-NUT center is
\be
B_{0}  = -  \mathbb{F}_0 + \frac{1}{16}\sum_{i=1}^{N}\frac{J_i}{R_i} \mathbb{F}_{p^0}.
\ee
The $B_j$ matrices anti-commute with each other, $\{ B_j, B_k\}=0$. It then follows that
the residues matrices $A_j$ are 
\bea
A_{j}  &=& - B_j  + \frac{1}{2 R_j} \{B_j,B_0 \} \\
&=& - B_j   - \frac{J_j}{16R_j} ( \mathbf{e}_{53} + \mathbf{e}_{71}) - \frac{Q_1^j}{4R_j} \mathbf{e}_{73},
\eea
and $A_0$ is 
\bea
A_{0}     &=& - B_0 -\frac{1}{2R_i} \sum_{i=1}^N \{B_0,B_i\} ,         \\
&=& - B_0 + \sum_{i=1}^{N} \frac{J_j}{16R_j} ( \mathbf{e}_{53} + \mathbf{e}_{71}) +\sum_{i=1}^{N} \frac{Q_1^j}{4R_j} \mathbf{e}_{73},
\eea
where symbol $\mathbf{e}_{ij}$ denote a $8 \times 8$ matrix with 1 in the $i$-th row and $j$-th column and 0 elsewhere.

\subsection*{Spectral flowed geometries with multiple supertubes}
An $N+1$-center bubbling solution associated to the above configuration  can be obtained by applying a combination of gauge and spectral flow transformations \cite{Bena:2008wt}. The final set of harmonic functions are,
\begin{align}
V   &= \e_0(1+\gamma c) + \gamma k^3_0 + \frac{1+\gamma\,c}{r} - \sum_{i=1}^N\frac{\gamma\,q_3^i}{2r_i}  ,       &
K^1  & = - \gamma\,l^2_0 - \sum_{i=1}^N\frac{\gamma\,Q_2^i}{4r_i},                                                          \nn         \\
K^2  & = - \gamma\,l^1_0 - \sum_{i=1}^N\frac{\gamma\,Q_1^i}{4r_i},                                                      &
K^3  & = k_0^3 + c\,\e_0 + \frac{c}{r} - \sum_{i=1}^{N}\frac{q_3^i}{2r_i}   ,                                         \nn         \\
M  & = m_0 - \frac{c\,l^3_0}2 - \sum_{i=1}^{N}\frac{J_i}{16r_i},                                                      &
L_1  & = l_0^1 + \sum_{i=1}^{N}\frac{Q_1^i}{4r_i},                                                                          \nn         \\
L_2  & = l_0^2 + \sum_{i=1}^{N}\frac{Q_2^i}{4r_i},                                                                      &
L_3  & = l^3_0 - 2\gamma\,m_0 + \gamma\,c\,l^3_0 + \sum_{i=1}^N\frac{\gamma\,J_i}{8r_i}               ,                 \nn
\end{align}
where 
$c = \frac{1}{2}\sum_{i=1}^{N}q_3^i$ so that the net Gibbons-Hawking charge is +1. Smoothness conditions \eqref{smoothness_eqs} and \eqref{bubbling_eqs} remain the same.
The  $B_j$ and $B_0$ matrices are
\bea
B_j             &=&\ \frac{\gamma\, q_3^j}2 \mathbb{F}_0 - \frac{Q_1^j}4 \mathbb{F}_1 - \frac{Q_2^j}4 \mathbb{F}_2 - \frac{\gamma J_j}8 \mathbb{F}_3 - \frac{\gamma\,Q^j_2}4 \mathbb{F}_{p^1} \nn \\
 & &- \frac{\gamma\,Q_1^j}4 \mathbb{F}_{p^2} - \frac{q_3^j}2 \mathbb{F}_{p^3}  + \frac{J_j}8 \mathbb{E}_{q^0}
 - \frac{J_j}{16R_j} \mathbb{F}_{p^0}     ,\\
B_{0}         &=&\ -(1+\gamma\,c) \mathbb{F}_0 + m_0 \mathbb{F}_{p^0} + c \mathbb{F}_{p^3}.                
\eea
The $B_j$ matrices anti-commute with each other, $\{ B_j, B_k\}=0$. It then follows that
the residues matrices $A_j$ are 
\bea
A_{j}  &=& - B_j  + \frac{1}{2 R_j} \{B_j,B_0 \} \\
&=& - B_j   - \frac{J_j}{16R_j} ( \mathbf{e}_{53} + \mathbf{e}_{71}) - \frac{Q_1^j}{4R_j} \mathbf{e}_{73},
\eea
and $A_0$ is 
\bea
A_{0}     &=& - B_0 -\frac{1}{2R_i} \sum_{i=1}^N \{B_0,B_i\} ,         \\
&=& - B_0 + \sum_{i=1}^{N} \frac{J_j}{16R_j} ( \mathbf{e}_{53} + \mathbf{e}_{71}) +\sum_{i=1}^{N} \frac{Q_1^j}{4R_j} \mathbf{e}_{73}.
\eea

\section{Discussion and future directions }

\label{sec:disc}

In vacuum five-dimensional gravity, many of the known interesting solutions  can be constructed via the inverse scattering approach.    The approach developed in \cite{KKV1, KKV2, KKV3,Chakrabarty:2014ora} for five-dimensional STU supergravity is akin to the inverse scattering approach of vacuum gravity. It has been shown in those references that certain non-extermal black holes  and certain non-supersymmetric bubbling solutions of STU supergravity are captured in that formalism. 

One of the main motivation of the present work is to extend that discussion to collinear BPS solutions. In this paper,  we have made significant progress on this problem, though some important questions remain unexplored. We obtained the Geroch group matrices for collinear bubbling solution, and exhibited that for these solutions the SO(4,4) monodromy matrices have only simple poles in the spectral parameter $w$ with residues of rank two. These are precisely the conditions under which the formalism developed in \cite{KKV2} is applicable. In this work, we have not attempted explicit factorisation of these matrices, though we have checked that various consistency conditions of \cite{KKV2} are satisfied.

The Geroch group description  obtained in this paper is valuable for future developments. As mentioned in the introduction, to find explicit (novel) solutions in the Riemann-Hilbert approach one needs to perform a  canonical factorisation of a monodromy matrix. Which monodromy matrix should one pick to start with? One strategy to obtain novel solutions can be to modify monodromy matrices of the known solutions in some controlled systematic way. This is the strategy we hope  to adopt in the future.  We would like to modify the monodromy matrices of bubbling solutions with $N$ centers so as to make $2k$ of its centers akin to that of the JMaRT solution. Then, naively it appears that upon factorisation one would be able to construct a solution with $N-2k$ BPS centers with  $k$ JMaRT type bolts.  

It will be interesting to explore the Geroch group description of the full class of BPS, non-BPS, and almost BPS solutions and better understand  the double (or higher) pole structure of monodromy matrices.  We except that the techniques of \cite{Camara:2017hez, Cardoso:2017cgi} to be applicable for their explicit factorisation. It will also be interesting to relate monodromy matrices to rod-structure \cite{Breunholder:2017ubu} in some precise way. 

We hope to return to some of these problems in our future work.

\subsection*{Acknowledgements} It is a pleasure to thank Guillaume Bossard and Axel Kleinschmidt for discussions. We thank AEI Potsdam for warm hospitality at the beginning stages of this project. Our work is supported in part  by the DST-Max Planck Partner Group ``Quantum Black Holes'' between IOP Bhubaneswar/CMI Chennai and AEI Potsdam. A preliminary version of these results were presented at the ``National Strings Meeting 2017'' at NISER Bhubaneswar in December 2017. We thank the audience for useful feedback. We are grateful to Axel Kleinschmidt for reading multiple versions of the draft and suggesting improvements. 

\appendix

\section{Explicit representatives for smaller nilpotent orbits of $\mf{so}$(4,4) Lie algebra}
\label{app:nilpotent_orbits}
Let us recall that a linear transformation $X$ on a finite dimensional vector space is called nilpotent if for some $r$, $X^r =0$. 
Let $\mf{g}_\mathbb{C}$ be an arbitrary complex Lie algebra. Let $\mf{h}_{\mathbb{C}}$ denote its Cartan subalgebra. We say an element $X$ of the complex Lie algebra $\mf{g}_\mathbb{C}$  is nilpotent if $\mbox{ad}_X$  is nilpotent. A classification of nilpotent elements of the Lie algebra is obtained through the study of conjugacy classes of such elements using the natural action of Lie group attached to the Lie algebra. These conjugacy classes are called nilpotent orbits. For applications to black holes in a theory with three-dimensional symmetry $G/K$, we need to know the $K$ orbits of nilpotent elements of the Lie algebra of $G$. Classifying all such orbits is a detailed exercise, see e.g., \cite{CollingwoodMcGovern, Dietrich:2016ojx, Ruggeri:2016dfk}. In this appendix we are interested in listing representatives of the some of the smaller orbits of the real Lie algebra $\mf{so}(4,4)$. We exhibit these representatives in the vector representation. We are interested in knowing ranks and nilpotency degree of these representatives in the vector representation.
This data is of interest in the main text of our paper.  

\subsection*{$\mf{so}(8,\mathbb{C})$ basis}

In order to list the representatives we define a basis. We do this in steps. Let us define four vectors in a four dimensional vector space, 
\begin{align}
v_1 &= \{1,0,0,0\}, &
v_2 &= \{0,1,0,0\},\\
v_3 &= \{0,0,1,0\},&
v_4 &= \{0,0,0,1\}, 
\end{align}
then the twelve positive roots for the $\mf{so}(8, \mathbb{C})$ Lie algebra can be taken to be
\begin{align}
& e_1 = v_1 - v_2,  &  
& e_2 =  v_1 - v_3,  & 
& e_3 = v_1 - v_4, &\\ 
& e_4 =  v_2 - v_3, &  
& e_5 = v_2 - v_4,  & 
& e_6 =  v_3 - v_4,  & \\
& e_7 =  v_1 + v_2,  &  
& e_8 = v_1 + v_3, &  
& e_9 = v_1 + v_4, & \\
& e_{10} = v_2 + v_3, &  
& e_{11} =  v_2 + v_4,  &   
& e_{12} = v_3 + v_4. &
\end{align}
The $8 \times 8$ dimensional Cartan-Weyl basis that realises the above root vectors is as follows. Let the symbol $\mathbf{e}_{ij}$ denote a $8 \times
8$ matrix with 1 in the $i$-th row and $j$-th column and 0 elsewhere. We have the Cartan subalgebra generators
\begin{align}
& \mathbb{H}_1 = \mathbf{e}_{11} - \mathbf{e}_{55}, &
& \mathbb{H}_2 = \mathbf{e}_{22} - \mathbf{e}_{66},\nn \\
& \mathbb{H}_3 = \mathbf{e}_{33}  - \mathbf{e}_{77} ,&
& \mathbb{H}_4 = \mathbf{e}_{44} -  \mathbf{e}_{88}, \label{basis1}
\end{align}
and the positive root generators
\begin{align}
& \mathbb{E}_1 = \mathbf{e}_{12} - \mathbf{e}_{65},  &  
&  \mathbb{E}_2 = \mathbf{e}_{13} - \mathbf{e}_{75},  & 
&  \mathbb{E}_3 = \mathbf{e}_{14} -\mathbf{e}_{85}, &\\ 
&  \mathbb{E}_4 = \mathbf{e}_{23} - \mathbf{e}_{76}, &  
&  \mathbb{E}_5 = \mathbf{e}_{24} - \mathbf{e}_{86}, &  
&  \mathbb{E}_6 = \mathbf{e}_{34} - \mathbf{e}_{87},  & \\
&  \mathbb{E}_7 = \mathbf{e}_{16} - \mathbf{e}_{25},  &  
&  \mathbb{E}_8 = \mathbf{e}_{17} - \mathbf{e}_{35},  & 
&  \mathbb{E}_9 = \mathbf{e}_{18} - \mathbf{e}_{45}, & \\
&  \mathbb{E}_{10} =  \mathbf{e}_{27} - \mathbf{e}_{36}, &  
&  \mathbb{E}_{11} =  \mathbf{e}_{28} - \mathbf{e}_{46}, &  
&  \mathbb{E}_{12} =  \mathbf{e}_{38} - \mathbf{e}_{47}, & \label{basis2}
\end{align}
together with the negative root generators,
\be
 \mathbb{F}_i =  \mathbb{E}_i^T.\label{basis3}
\ee
The simple step generators are $\mathbb{E}_1, \mathbb{E}_2, \mathbb{E}_3$, and $\mathbb{E}_{12}$. 

\subsection*{$\alpha$- and $\beta$-labels}

For each nilpotent orbit of the complex Lie algebra 
$\mf{g}_\mathbb{C}$
there is a triple $(e, f, h)$ of elements where $e$ is a nilpotent element of the orbit, such that~\cite{CollingwoodMcGovern}
\begin{align}
[h,e] &= 2 e, & [h,f] &= - 2 f, & [e,f]&=h .
\end{align}
Such a triple is called a standard triple. We can take $h$ to be in the Cartan subalgebra $\mf{h}_{\mathbb{C}}$, and furthermore conjugate $\{h, e,f \}$ such that $h$  lies in the ``fundamental domain'' \cite{CollingwoodMcGovern}. Then the element $h$ is characterized by its  eigenvalues under the adjoint action on simple root generators $e_i$'s
\be
[h, e_i] = \alpha_i (h) e_i, \qquad \qquad i = 1, 2, \ldots, \mbox{rank}~\mf{g}.
\ee
Moreover, we can always find a standard triple such that the eigenvalues $\alpha_i (h) \in \{0,1,2\}$. If we label every node of the Dynkin diagrams with the eigenvalue $\alpha_i (h)$ of $h$ on the corresponding simple root, we get the weighted Dynkin diagrams for the orbits. These are called $\alpha$-label of the nilpotent orbit. There are twelve such orbits\footnote{In the table on page 84 of \cite{CollingwoodMcGovern} there is a typo. Our $\mathcal{O}_6$ orbit in their notation corresponds to the partition [3$^2$, 1$^2$] and is not listed in the table.} for SO$(8, \mathbb{C})$.  We are only interested in the smaller orbits with weighted Dynkin diagrams:

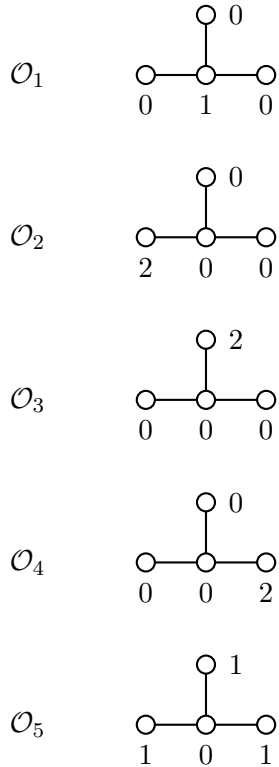
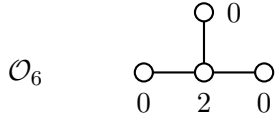
\begin{center}
  \begin{tikzpicture}[scale=.4]
    \draw (-1,0) node[anchor=east]  {$\mathcal{O}_1$};
    \foreach \x in {1,...,3}
    \draw[thick,xshift=\x cm] (\x cm,0) circle (3 mm);
    \foreach \y in {1,...,2}
    \draw[thick,xshift=\y cm] (\y cm,0) ++(.3 cm, 0) -- +(14 mm,0);
    \draw[thick] (4 cm,2 cm) circle  (3 mm);
    \draw[thick] (4 cm, 3mm) -- +(0, 1.4 cm);
    \node[]  at (2,-1) {0};
    \node[]  at (4,-1) {1};
    \node[]  at (6,-1) {0};
    \node[]  at (5,2) {0};
 \end{tikzpicture}
\end{center}

\begin{center}
  \begin{tikzpicture}[scale=.4]
      \draw (-1,0) node[anchor=east]  {$\mathcal{O}_2$};
  \foreach \x in {1,...,3}
    \draw[thick,xshift=\x cm] (\x cm,0) circle (3 mm);
    \foreach \y in {1,...,2}
    \draw[thick,xshift=\y cm] (\y cm,0) ++(.3 cm, 0) -- +(14 mm,0);
    \draw[thick] (4 cm,2 cm) circle  (3 mm);
    \draw[thick] (4 cm, 3mm) -- +(0, 1.4 cm);
    \node[]  at (2,-1) {2};
    \node[]  at (4,-1) {0};
    \node[]  at (6,-1) {0};
    \node[]  at (5,2) {0};
 \end{tikzpicture}
\end{center}

\begin{center}
  \begin{tikzpicture}[scale=.4]
    \draw (-1,0) node[anchor=east]  {$\mathcal{O}_3$};
    \foreach \x in {1,...,3}
    \draw[thick,xshift=\x cm] (\x cm,0) circle (3 mm);
    \foreach \y in {1,...,2}
    \draw[thick,xshift=\y cm] (\y cm,0) ++(.3 cm, 0) -- +(14 mm,0);
    \draw[thick] (4 cm,2 cm) circle  (3 mm);
    \draw[thick] (4 cm, 3mm) -- +(0, 1.4 cm);
    \node[]  at (2,-1) {0};
    \node[]  at (4,-1) {0};
    \node[]  at (6,-1) {0};
    \node[]  at (5,2) {2};
 \end{tikzpicture}
\end{center}

\begin{center}
  \begin{tikzpicture}[scale=.4]
    \draw (-1,0) node[anchor=east]  {$\mathcal{O}_4$};
    \foreach \x in {1,...,3}
    \draw[thick,xshift=\x cm] (\x cm,0) circle (3 mm);
    \foreach \y in {1,...,2}
    \draw[thick,xshift=\y cm] (\y cm,0) ++(.3 cm, 0) -- +(14 mm,0);
    \draw[thick] (4 cm,2 cm) circle  (3 mm);
    \draw[thick] (4 cm, 3mm) -- +(0, 1.4 cm);
    \node[]  at (2,-1) {0};
    \node[]  at (4,-1) {0};
    \node[]  at (6,-1) {2};
    \node[]  at (5,2) {0};
 \end{tikzpicture}
\end{center}

\begin{center}
  \begin{tikzpicture}[scale=.4]
    \draw (-1,0) node[anchor=east]  {$\mathcal{O}_5$};
    \foreach \x in {1,...,3}
    \draw[thick,xshift=\x cm] (\x cm,0) circle (3 mm);
    \foreach \y in {1,...,2}
    \draw[thick,xshift=\y cm] (\y cm,0) ++(.3 cm, 0) -- +(14 mm,0);
    \draw[thick] (4 cm,2 cm) circle  (3 mm);
    \draw[thick] (4 cm, 3mm) -- +(0, 1.4 cm);
    \node[]  at (2,-1) {1};
    \node[]  at (4,-1) {0};
    \node[]  at (6,-1) {1};
    \node[]  at (5,2) {1};
 \end{tikzpicture}
\end{center}

\begin{center}
  \begin{tikzpicture}[scale=.4]
    \draw (-1,0) node[anchor=east]  {$\mathcal{O}_6$};
    \foreach \x in {1,...,3}
    \draw[thick,xshift=\x cm] (\x cm,0) circle (3 mm);
    \foreach \y in {1,...,2}
    \draw[thick,xshift=\y cm] (\y cm,0) ++(.3 cm, 0) -- +(14 mm,0);
    \draw[thick] (4 cm,2 cm) circle  (3 mm);
    \draw[thick] (4 cm, 3mm) -- +(0, 1.4 cm);
    \node[]  at (2,-1) {0};
    \node[]  at (4,-1) {2};
    \node[]  at (6,-1) {0};
    \node[]  at (5,2) {0};
 \end{tikzpicture}
\end{center}
where we use the standard ordering convention
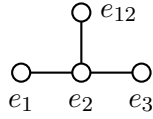
\begin{center}
  \begin{tikzpicture}[scale=.4]
    \foreach \x in {1,...,3}
    \draw[thick,xshift=\x cm] (\x cm,0) circle (3 mm);
    \foreach \y in {1,...,2}
    \draw[thick,xshift=\y cm] (\y cm,0) ++(.3 cm, 0) -- +(14 mm,0);
    \draw[thick] (4 cm,2 cm) circle  (3 mm);
    \draw[thick] (4 cm, 3mm) -- +(0, 1.4 cm);
    \node[]  at (2,-1) {$e_1$};
    \node[]  at (4,-1) {$e_2$};
    \node[]  at (6,-1) {$e_3$};
    \node[]  at (5.25,2) {$e_{12}$};
 \end{tikzpicture}
\end{center}
with $\{e_1, e_2, e_3, e_{12}\}$ as simple root generators.

The $\alpha$-labels uniquely specify the nilpotent orbits of the complex Lie algebra $\mf{so}(8, \mathbb{C})$.  It may happen that for two nilpotent elements $X, Y$ in  $\mf{so}(4,4)$ are in the same $\mf{so}(8, \mathbb{C})$ orbit but different $\mf{so}(4,4)$ orbits, i.e., 
\be
g X g^{-1} = Y,
\ee
for some $g$ in SO$(8, \mathbb{C})$ but not for any $g$ in SO(4,4). That is,  for a given $\alpha$-label there can be several real orbits. The real orbits are classified by the $\beta$-labels. 
The $\beta$-labels are defined as follows. For an SO(8,$\mathbb{C}$) orbit consider a standard triple $(h,e,f)$ such that under Chevalley involution $\theta$,
\be
\theta (h) = -h, \quad \theta(e) = -f, \quad \theta(f) = -e.
\ee
Such standard triples are called Cayley triples. All the triples listed below for SO(8,$\mathbb{C}$) orbits are of this type. The Cayley transform of a Cayley triple $(e,f,h)$ is now defined as the standard triple
$(e',f',h')$ where
\bea
e' &=& \frac{1}{2} (e + f + i h), \\
f' &=& \frac{1}{2} (e + f - i h), \\
h' &=& i (e - f).
\eea
It follows that $h' \in \mathfrak{k}_\mathbb{C}$ and $e', f' \in \mathfrak{p}_\mathbb{C}$,  where
$\mathfrak{k}_\mathbb{C}$ is the subalgebra of $\mathfrak{g}_\mathbb{C}$ fixed under the Chevalley involution.
The algebra $\mathfrak{k}_\mathbb{C}$ in our case is 
\be
\left[ \mathfrak{sl}(2, \mathbb{C}) \right]^4.
\ee
One can take the $h'$ in the Cartan subalgebra of $\mathfrak{k}_\mathbb{C}$. We label each node of the Dynkin diagram of $\mathfrak{k}_\mathbb{C}$ by the eigenvalue of the corresponding simple root on $h'$. The weighted Dynkin diagram thus obtained is called the $\beta$-label of the orbit, and is an invariant of the real orbits of SO(4,4). $\alpha$- and $\beta$- labels are classified  for the case of our interest in \cite{Dietrich:2016ojx}.

\subsection*{$\left[ \mathfrak{sl}(2, \mathbb{C}) \right]^4$ generators}
There are many ways to make manifest the four commuting $\mathfrak{sl}(2, \mathbb{C})$. We find a set that will be useful for the smaller orbits, in the sense that $h'$ will be in Cartan subalgebra of $\mathfrak{k}_\mathbb{C}$.

Recall that we have 12 positive roots $e_i$'s and 12 negative roots $f_i$'s. The set of generators fixed under the Chevalley involution are of the type
\be
k_i = e_i - f_i.
\ee

We start with observation that $k_1, k_6, k_7$ and $k_{12}$ commute with each other. These are a maximal set of commuting generators, so they can be taken to span the Cartan subalgebra of  $\mathfrak{k}_\mathbb{C}$.
Consider the general element in the Cartan subalgebra 
\be
X = a_1 k_1 + a_6 k_6  + a_7 k_7 + a_{12} k_{12},
\ee
and consider $\mathrm{ad}_{X} k_i$, $i=1,2, \ldots, 12$, i.e.,
\be
[X,k_i] = \sum_{j=1}^{12}c^j_i k_j.
\ee
We observe that by tuning $(a_1,a_6,a_7,a_{12})$ we can make 10 out of 12 eigenvalues of the matrix $c^j_i$ zeros. The combinations that diagonalise the matrix $c^j_i$ make the four commuting
$\mathfrak{sl}(2, \mathbb{C})$ manifest. Such a set of generators are
\bea
H_1 &=&  \frac{i}{2} \left( k_1 + k_6 + k_7 + k_{12}\right), \\
H_2 &=& \frac{i}{2} \left( k_1 - k_6 + k_7 - k_{12}\right),\\
H_3 &=& \frac{i}{2} \left( k_1 + k_6 - k_7 - k_{12}\right), \\
H_4 &=& \frac{i}{2} \left( k_1 - k_6 - k_7 + k_{12}\right),
\eea

\bea
E_1 &=&  \frac{i}{4} \left( -k_2 + i k_3 + i k_4  + k_5 -k_8 + i k_9 + i k_{10}  + k_{11}\right), \\
E_2 &=&  \frac{i}{4} \left(+ k_2 + i k_3 - i k_4  + k_5 + k_8 + i k_9 - i k_{10}  + k_{11}\right), \\
E_3 &=&  \frac{i}{4} \left( +k_2 - i k_3 - i k_4  - k_5 -k_8 + i k_9 + i k_{10}  + k_{11}\right), \\
E_4 &=&  \frac{i}{4} \left( -k_2 - i k_3 + i k_4  - k_5 +k_8 + i k_9 - i k_{10}  + k_{11}\right),
\eea
\bea
F_1 &=&  \frac{i}{4} \left( -k_2 - i k_3 - i k_4  + k_5 -k_8 - i k_9 - i k_{10}  + k_{11}\right), \\
F_2 &=&  \frac{i}{4} \left( +k_2 - i k_3 + i k_4  + k_5 +k_8 - i k_9 + i k_{10}  + k_{11}\right), \\
F_3 &=&  \frac{i}{4} \left( +k_2 + i k_3 + i k_4  - k_5 -k_8 - i k_9 - i k_{10}  + k_{11}\right), \\
F_4 &=&  \frac{i}{4} \left( -k_2 + i k_3 - i k_4  - k_5 +k_8 - i k_9 + i k_{10}  + k_{11}\right).
\eea

\subsection*{$\mathcal{O}_1$ orbit}
As an example consider the $\mathcal{O}_1$ orbit. A standard triple for this orbit is\footnote{To avoid any confusion, let us remark that the $h$ is in the Cartan subalgebra, but is not in the fundamental domain. Hence, its eigenvalues  on the simple root generators do not give the $\alpha$-label. The same comment applies to the standard triples listed in table \ref{table}.}
\bea
h &=& h_1 - h_2, \\
e&=&e_1, \\
f &=& f_1.
\eea
In the vector representation, matrices in this orbit have nilpotency $X^2 =0$ and have $\verb+rank+ \ X = 2$. The $\alpha$ label for this orbit is 
\be
(0,1,0,0),
\ee
Now using Cayley transform 
\be
h' = i (e_1-f_1) = i k_1 = \frac{1}{2}(H_1 + H_2 + H_3 + H_4).
\ee
Therefore the $\beta$-label is  $(1,1,1,1)$. The $\mathcal{O}_1$ orbit is of main interest in the main text of our paper. It is the unique orbit with nilpotency degree two in the vector representation, and matrix rank of representatives in this orbit is 2. In the adjoint this orbit has degree 3, i.e., 
\be
(\mathrm{ad}_{X})^3=0.
\ee
With a bit more work, one can similarly find other representatives. On the next page we present a list.

\newpage

 \newcolumntype{Q}{>{\centering\arraybackslash}X}
 \newcolumntype{Y}{>{\raggedright\arraybackslash}X}
  \newcolumntype{Z}{>{\centering\arraybackslash$\displaystyle}X<{$}}
\renewcommand{\tabularxcolumn}[1]{>{\arraybackslash}m{#1}}
\begin{table}[h!]
\begin{center}
\begin{tabularx}{\textwidth}{p{1.2cm}p{1.6cm}Zp{1.7cm}Zp{.9cm}p{1.1cm}}
\toprule%
\textbf{Orbit}  &   $\a$\textbf{-label} & \textbf{Neutral} &   $\b$\textbf{-label}  & \textbf{Nilpositive}   &   \textbf{Nilp.}  &   \textbf{Rank($X$)}     \\  \midrule
$[2^2,1^4]$ & $(0,1,0,0)$   & h_1 - h_2    & $(1,1,1,1)$   & e_1   & 2  & 2 \\  \midrule    
\multirow{2}*{$[3,1^5]$}   & \multirow{2}*{$(2,0,0,0)$}  & \multirow{2}*{$2h_1$}  & $(2,2,0,0)$   & e_1+e_7   &   \multirow{2}*{$3$} &     \multirow{2}*{$2$} \\  
    &   &  & $(0,0,2,2)$   & e_1-e_7   &    &         \\  \midrule
\multirow{2}*{$[2^4]^I$}    & \multirow{2}*{$(0,0,2,0)$}   &  \multirow{2}*{$h_1-h_2+h_3-h_4$}     & $(2,0,2,0)$ & e_1+e_6    &   \multirow{2}*{$2$} &  \multirow{2}*{$4$}    \\ 
    &   &   &  $(0,2,0,2)$     & e_1-e_6 &   &   \\ \midrule
\multirow{2}*{$[2^4]^{II}$}    &   \multirow{2}*{$(0,0,0,2)$}   &   \multirow{2}*{$h_1-h_2+h_3+h_4$}  &   $(2,0,0,2)$ &   e_1+e_{12}  &   \multirow{2}*{$2$} &   \multirow{2}*{$4$} \\ 
    &   &   &   $(0,2,2,0)$ &   e_1-e_{12}  &  & \\  \midrule
\multirow{4}*{$[3,2^2,1]$}    &   \multirow{4}*{$(1,0,1,1)$}   &   2h_1 + h_3 -h_4  &   $(3,1,1,1)$ &   e_1+e_6+e_7  &   \multirow{4}*{$3$} &   \multirow{4}*{$4$} \\ 
    &   &  2h_1 + h_3 +h_4 &   $(1,3,1,1)$ &   e_1+e_7-e_{12}  &   &   \\  
    &   &  h_1 -h_2+ 2h_3  &   $(1,1,3,1)$ &   e_1+e_6-e_{12}  &   &   \\  
    &   &  h_1 +h_2 + 2h_3 &   $(1,1,1,3)$ &   e_6+e_7-e_{12}  &   &   \\  \midrule
\multirow{5}*{$[3^2,1^2]$}    &   \multirow{5}*{$(0,2,0,0)$}   &   \multirow{4}*{$2(h_1+h_3)$}  &   $(4,0,0,0)$ &   e_1+e_6+e_7+e_{12}  &   \multirow{5}*{$3$} &   \multirow{5}*{$4$} \\ 
    &   &   &   $(0,4,0,0)$ &   e_1-e_6+e_7-e_{12}  &   &   \\  
    &   &   &   $(0,0,4,0)$ &   e_1+e_6-e_7-e_{12}  &   &   \\  
    &   &   &   $(0,0,0,4)$ &   e_1-e_6-e_7+e_{12}  &   &   \\  \cmidrule(l){3-5}
    &   &   2(h_1-h_3)  &   $(2,2,2,2)$ &   \sqrt{2}(e_1+e_4)  &   &   \\  
\bottomrule%
\end{tabularx}  
\caption{List of the smaller nilpotent orbits of $\so(4,4)$ classified by $\a,\b$-labels. We have also listed representative neutral and nilpositive elements that together form a standard triple $\{h,e,f\}$ in the corresponding orbit. Nilp. and Rank$(X)$ denote the nilpotency degree (smallest $n$ such that $X^n=0$) and the matrix rank of the orbit in the 8 $\times$ 8 vector representation.}
\label{table}
\end{center}
\end{table}

\end{document}